\documentclass[11pt]{article}
\usepackage[english]{babel}
\usepackage{natbib}
\usepackage{comment}
\usepackage{float}
\usepackage{mathrsfs}
\usepackage{enumitem}
\usepackage[font={small,it}]{caption}
\usepackage{amsmath,amsfonts,amsthm,amssymb}
\usepackage{bm,rotating,dsfont,graphicx}
\usepackage{url}
\usepackage{multicol}
\usepackage{multirow}
\usepackage[T1]{fontenc}
\usepackage{flafter}
\usepackage{appendix}
\usepackage{subfigure}
\usepackage{xcolor}
\usepackage{soul}
\usepackage{setspace}
\usepackage{booktabs}
\usepackage{tabularx}
\usepackage{threeparttable}
\usepackage[hidelinks]{hyperref}
\makeatletter
\def\hlinewd#1{%
	\noalign{\ifnum0=`}\fi\hrule \@height #1 %
	\futurelet\reserved@a\@xhline}
\makeatother
\def\spacingset#1{\renewcommand{\baselinestretch}{#1}\small\normalsize}
\spacingset{1}

\begin{document}

\title{\textbf{Network Dynamics and Spatial Shifts in Civilian Targeting: A Stochastic Block Model Analysis of the Colombian Armed Conflict}}

\date{}

\author{
    Natalia Perdomo-Londoño\qquad\qquad Juan Sosa\footnote{Corresponding author: jcsosam@unal.edu.co.}\qquad\qquad Emma Camargo-Díaz \\
    \hspace{1cm}\\
    {\small Universidad Nacional de Colombia, Colombia} \\
    {\small Universidad Externado de Colombia, Colombia}
}

\maketitle

\begin{abstract}
In this article, we explore how the escalating victimization of civilians during civil wars is mirrored in the fragmented distribution of territorial control, focusing on the Colombian armed conflict. Through an exhaustive characterization of the topology of bipartite and projected networks of municipalities, we describe changes in territorial configurations across different periods between 1978 and 2007. By employing stochastic block models for count data, we show that, during periods dominated by a small set of actors, the networks adopt a centralized node periphery structure, whereas during times of widespread conflict, areas of influence overlap in complex ways. Our findings also suggest the existence of cohesive municipal communities shaped by both geographic proximity and affinities between armed structures, as well as internally dispersed groups with a high likelihood of interaction. As the spatial distribution shifts toward a more fragmented arrangement, the average interaction intensity between communities predicted by the stochastic block model approaches that within communities, indicating a weakening of modular structure and increased inter community connectivity.
\end{abstract}

\noindent
{\it Keywords: Armed conflict; Bipartite networks; Community detection;   Social networks; Stochastic block model; Territorial dispute.}

\spacingset{1.1} 

\newpage

\section{Introduction}

During internal armed conflicts or civil wars, sovereignty becomes fragmented as the use of force is no longer monopolized by the State. Various actors compete for territorial control not only through direct military confrontations but also by targeting noncombatants, leading to territorial disputes that disproportionately affect civilians. Because irregular armies often face constraints in sustaining prolonged warfare, they may resort to violence against civilian populations, among other strategies, to consolidate regional power \citep[][p.87]{Kalyvas2006}.

Population control is crucial for armed groups, since communities provide key resources such as shelter, supplies, and information. An armed group’s capacity to challenge rivals for control and to impose an alternative order depends largely on the effectiveness of its strategies of subjugation over the population. Through intimidation, these groups secure obedience, punish infractions, and deter collaboration with adversaries. Accordingly, the type and intensity of violence they employ become instruments to establish and maintain control \citep{Arjona2016}.

\cite{Kalyvas2006} examines how attacks on civilians vary with the degree of territorial control exercised by an armed actor. When control is limited or contested, violence tends to be indiscriminate and extensive, as armed groups use it to secure their position and neutralize potential support for the enemy. By contrast, under exclusive control, selective violence is more effective and is typically associated with substantially lower levels of victimization. In such settings, exploiting the resources that the population can provide, including recruitment through forced enlistment, becomes more advantageous than relying on widespread coercion.

According to Kalyvas, civil wars are marked by high levels of territorial fragmentation during peaks of armed confrontation. Armed groups form ``corridors'' and induce relationships between noncontiguous regions by expanding their spheres of influence, establishing cooperative or competitive ties with other groups, or generating divisions that segment space. Rather than being fixed or absent, the boundaries separating locations in irregular war are diffuse and fluid. When disputes are intense, fragmentation becomes apparent as actors intermittently converge across multiple locations. Irregular warfare thus reshapes sovereignty, and the resulting spatial and territorial fragmentation reflects this shift. By contrast, when an actor consolidates territorial control, clearly delineated zones of influence tend to emerge.

Graphs provide a natural representation of complex relationships between armed actors and regions during an armed conflict. This perspective supports the visualization of how interactions are configured and evolve within fragmented territories, where presence and control vary over time with the dynamics of the conflict. Statistical network analysis \citep[e.g.,][]{menzel2010introduction,KolaczykCsardi2014,Luke2015} offers suitable tools for describing these relationships. In particular, stochastic block models \citep{holland1983stochastic,nowicki2001estimation,KarrerNewman2010,aicher2014learning} can reveal fragmentation or segmentation by identifying communities of nodes with similar connectivity patterns.

In Colombia, a salient feature of recent internal armed conflicts is the systematic victimization of civilians as a strategy for contesting territorial control, as documented by the \textit{Centro Nacional de Memoria Histórica} \citep{CNMH2013}. This study uses network analysis to examine how the relationships between the most prominent irregular armed groups and the municipalities they occupied evolved between 1978 and 2007. In addition, by employing stochastic block models for count data, we assess whether municipalities within the zones of influence of these armed actors formed clearly differentiated communities or whether these territories were highly interconnected, consistent with the fragmented and fluid nature of the conflict.

Based on information collected by the CNMH on the most impactful acts of victimization against noncombatants, including massacres, selective assassinations, forced disappearances, kidnappings, and forced recruitment, we define ten bipartite networks linking municipalities and armed structures between 1978 and 2007, together with their corresponding projections. Our analysis focuses on the most prominent irregular armies in the conflict, the \textit{Fuerzas Armadas Revolucionarias de Colombia} (FARC) and the \textit{Autodefensas Unidas de Colombia} (AUC). Although the AUC existed nominally from 1997 to 2005, it had clear antecedents in the early 1980s, so we include the paramilitary structures that preceded the AUC.

The paper is organized as follows. Section 2 introduces the graph framework, the municipality projections, and stochastic block models for weighted networks. Section 3 describes the data sources, periodization, and the construction of the bipartite incidence matrices and municipal projection networks, together with descriptive network features across periods. Section 4 fits Poisson stochastic block models to each projected network and summarizes the evolution of community structure over time. Section 5 concludes with a discussion of substantive implications and limitations.

\section{Methodology}\label{sec2}

In this section, we present the main methodological elements of community detection. We first provide a concise characterization of graphs and bipartite graphs, and then discuss clustering and stochastic block models for count data. For an accessible introduction, see \cite{menzel2010introduction}, \cite{KolaczykCsardi2014}, and \cite{Luke2015}, among others.

\subsection{Graphs and bipartite graphs}

A graph is a mathematical structure composed of vertices (nodes or actors) and edges (links or connections) that join pairs of vertices, and it is widely used to represent relational data. Formally, a graph $G$ is given by \( G = (V, E) \), where \( V \) denotes the set of vertices and \( E \) denotes the set of edges. Each edge is defined as a pair of vertices \( e = (u, v) \), with \( u, v \in V \).

A graph can be undirected, where \( (u, v) \in E \) implies \( (v, u) \in E \), so edges have no direction. It can also be directed, where \( (u, v) \in E \) does not necessarily imply \( (v, u) \in E \), so edges have a specific direction. The graph is called simple if it has no loops, meaning edges that connect a vertex to itself, and no multiple edges between the same pair of vertices. A weighted graph assigns a value to each edge through a function \( w: E \rightarrow \mathbb{R} \), for example counts, costs, or distances. An unweighted graph has no edge weights and represents only whether a connection between vertices is present or absent.

A bipartite graph is a type of graph where the vertex set can be partitioned into two disjoint sets, usually denoted by $U$ and $V$, such that every edge connects a vertex in $U$ to a vertex in $V$. Therefore, no edge links two vertices within the same set. This property can be characterized using coloring. A graph is bipartite if and only if it is 2 colored, meaning that its vertices can be colored with two colors so that no two adjacent vertices share the same color. Bipartite graphs are useful for modeling relationships between two distinct types of objects, for example users and products. For a detailed introduction to graph theory and related topics, see \cite{bondy1976graph}, \cite{west2001introduction}, and \cite{diestel2017graph}, among other comprehensive sources.

For each period $t$, we construct incidence matrices $\mathbf{A}_t=[A_{i,j,t}]$ that define bipartite graphs. The rows correspond to municipalities $\mathcal{M}$ and the columns correspond to armed structures $\mathcal{E}$. Entries are binary, with $A_{i,j,t}=1$ if armed substructure $e_j\in\mathcal{E}$ carried out civilian targeting in municipality $m_i\in\mathcal{M}$ during period $t$, and $A_{i,j,t}=0$ otherwise.

When projecting onto municipalities, we obtain an undirected weighted graph \(G^M_t\) using \(\mathbf{A}_t \mathbf{A}_t^{\textsf{T}}\) for each \(t\). In this graph, two municipalities \(m_i, m_j \in M\) are connected if and only if both were sites of activity by at least one common armed structure during the period. Connection intensity reflects the number of shared groups. In particular, each entry of the projected adjacency matrix \(\mathbf{Y}^\mathcal{M}_t = [Y^\mathcal{M}_{i,j,t}]\) equals the number of substructures that operated in both \(m_i\) and \(m_j\) during period \(t\).

The municipal projection $G_t^{\mathcal{M}}$ is the object of interest in this paper because it directly encodes territorial overlap among municipalities. In substantive terms, an edge between two municipalities indicates co-targeting, meaning that at least one armed substructure engaged in civilian targeting in both municipalities during the same period $t$, and the corresponding edge weight measures the extent of overlap through the number of shared substructures. This projection allows us to study how zones of influence and competition are organized across space, and how these overlap patterns evolve over time. At the same time, projecting necessarily discards information about which specific structures generate each tie and it mechanically induces complete subgraphs among municipalities that share an armed structure, which can inflate transitivity and create clique like artifacts. We accept this tradeoff because our estimand is explicitly municipality municipality overlap, not the reconstruction of the full bipartite interaction system. Bipartite stochastic block models and related latent block models are natural alternatives when the bipartite structure itself is the target of inference, but here the projection provides a direct representation of territorial overlap that can be modeled and compared across periods.

\subsection{Stochastic Block Models}

A stochastic block model (SBM) is a random graph framework for inferring groups or communities in a network. Following \cite{KolaczykCsardi2014}, consider a set of $N$ individuals, and let \(G = (V, E)\) denote the graph representing their relationships. Here, \(V\) is the set of nodes and \(E\) is the set of edges, given by the unordered pairs \((i, j)\) of distinct vertices \(i, j \in V\).

In an SBM, each vertex \(i \in V\) belongs to exactly one of \(Q\) disjoint and exhaustive classes, also referred to as groups, clusters, or communities, denoted by \(\textsf{C}_1,\ldots,\textsf{C}_Q\). Let \(\Lambda\) be a \(Q \times Q\) matrix whose entries \(\Lambda_{q,r}\) represent interaction intensities between vertices in classes \(q\) and \(r\). This matrix determines the large-scale structure of the network. For instance, if \(\Lambda\) is diagonal, the network decomposes into disconnected components. If the off diagonal entries are positive and typically smaller than the diagonal ones, the network exhibits communities with dense within class connectivity and sparse between class connectivity \citep{KarrerNewman2010}. Other choices of \(\Lambda\) can generate alternative structures such as core periphery or hierarchical configurations. For an undirected graph, \(\Lambda_{q,r} = \Lambda_{r,q}\), and the adjacency matrix \(\mathbf{Y}\) is symmetric.

Class membership is commonly modeled as independent draws from a categorical distribution on \(\{1,\ldots,Q\}\). Let \(Z_i = (Z_{i,1},\ldots,Z_{i,Q})\) be the membership indicator for vertex \(i\), with \(Z_{i,q} = 1\) if \(i\) belongs to class \(q\) and \(Z_{i,q} = 0\) otherwise. Then \(Z_i\) has a categorical distribution with parameters \(\theta_1,\ldots,\theta_Q\), where \(\theta_q > 0\) and \(\sum_q \theta_q = 1\), so that \(\Pr(Z_{i,q} = 1 \mid \theta_q) = \theta_q\).

For networks with count valued weights, a Poisson SBM assumes that each entry \(Y_{i,j}\) is conditionally Poisson given the class labels of vertices \(i\) and \(j\). Specifically, \(Y_{i,j} \mid Z_{i,q} = 1, Z_{j,r} = 1 \sim \textsf{Poisson}(\Lambda_{q,r})\). Let \(L_{q,r}\) denote the total interaction count between classes \(q\) and \(r\), given by summing \(Y_{i,j}\) over all pairs with \(i \in \textsf{C}_q\) and \(j \in \textsf{C}_r\). Then \(L_{q,r} \sim \textsf{Poisson}(\Lambda_{q,r}\,n_{q,r})\), where \(n_{q,r} = n_q n_r\) for \(q < r\) and \(n_{q,q} = n_q(n_q - 1)/2\), with \(n_q\) the number of vertices in class \(q\). Under this representation, the distribution can be written as
\[
\Pr(Y = y \mid \Lambda)
= \prod_{q \leq r}
\frac{\exp\big(-\Lambda_{q,r}\,n_{q,r}\big)\big(\Lambda_{q,r}\,n_{q,r}\big)^{L_{q,r}}}{L_{q,r}!}\,.
\]

In our setting, edges represent counts, for example the number of shared structures between two municipalities. This formulation models the intensity of connections through shared attributes, rather than using only the presence or absence of a relation, and it provides a flexible tool for analyzing networks with count data.

\section{Network construction and processing}

Given that violence has taken multiple forms and levels of severity, this research focuses on the most serious human rights violations, specifically those related to the rights to life, physical integrity, liberty, and personal security \citep{RamaJudicial}. Accordingly, we use data on massacres, enforced disappearances, targeted killings, kidnappings, and illegal recruitment attributed to FARC substructures and to paramilitary groups that operated between 1978 and 2007.

Direct military engagements between armed actors are not considered, since they reflect different dimensions of armed group behavior in civil wars. According to \cite{Kalyvas2006}, confrontation among belligerents aims to defeat the adversary and involves military strategies and combat tactics directed primarily at armed opponents. By contrast, violence against civilians follows a different logic. It does not necessarily seek immediate military advantage, instead it aims to consolidate territorial control, undermine support for rival actors, and secure local resources. This violence is typically selective and intended to shape civilian behavior in line with the perpetrator’s interests, through intimidation, retaliation, or cooptation.

To reconstruct the territorial presence of FARC and paramilitary structures, we rely on event level data compiled by the \textit{Observatorio de Memoria y Conflicto} (OMC), an initiative of Colombia’s \textit{Centro Nacional de Memoria Histórica} (CNMH). The OMC systematically documents human rights violations, breaches of international humanitarian law, and other forms of political violence associated with the armed conflict. For each repertoire of violence, the dataset includes two relational tables. One records individual cases and another identifies victims. Both provide anonymized information on the year and municipality of occurrence and on the armed group held responsible. These data are used to construct bipartite networks linking municipalities and armed structures, from which we derive municipal projections for subsequent analysis.

The OMC documented the victimization of 318,826 individuals for the types of violence considered in this study. Of these, 217,629 have records for the variables used to attribute responsibility. We identify 85,166 individuals whose victimization can be attributed to armed structures of the FARC, paramilitary groups, and organized crime, excluding groups formed after the demobilization of the AUC. The information by event type is reported in Table \ref{tabla_victimas_violencia}.

\begin{table}[!htb]
\centering
\caption{Number of victims by type of violence.}
\label{tabla_victimas_violencia}
\scriptsize
\begin{tabular}{lcccccc}
\toprule
Number of & Selective & Forced disa- & \multirow{2}{*}{Massacres} & \multirow{2}{*}{Kidnappings} & Illegal re- & Sexual \\
victims   & killings  & ppearances   &                             &                               & cruitment  & violence \\
\midrule
Documented                    & 178,112 & 74,550 & 24,562 & 35,542 & 6,061 & 14,847 \\
\midrule
Attribution data available    & 120,238 & 42,241 & 20,875 & 30,384 & 5,635 & 11,156 \\
\midrule
Armed group and timeline      & 46,287  & 15,015 & 10,061 & 12,238 & 2,910 & 494 \\
\bottomrule
\end{tabular}
\medskip
\footnotesize\emph{Source: Elaborated by the authors with data from the CNMH.}
\end{table}

Moreover, the OMC dataset does not systematically report the specific military unit responsible for each event attributed to the FARC or paramilitary organizations. To address this limitation and ensure spatial and temporal consistency in the networks, we consulted additional sources to identify the armed structures active in each period and to reconstruct their geographic scope of influence over time. This reconstruction is based on the historiographic series produced by the CNMH \citep{CNMH2013,CNMH2017,CNMH2017b,CNMH2018a,CNMH2018b,CNMH2019a,CNMH2019b,CNMH2020a,CNMH2020b,CNMH2021a,CNMH2021b,CNMH2021c,CNMH2021d,CNMH2021e,CNMH2022a,CNMH2022b,CNMH2022c,CNMH2022d,CNMH2022e,CNMH2022f,CNMH2022g,CNMH2022h,CNMH2022j,CNMH2023a,CNMH2023b,CNMH2023c,CNMH2023d,CNMH2023e,CNMH2023f}, the Commission for the Clarification of Truth, Coexistence and Non-Repetition (CEV) \citep{CEV2022a,CEV2022b,CEV2022c,CEV2022d,CEV2022e,CEV2022f,CEV2022g,CEV2022h,CEV2022j}, and a systematic review of \textit{Noche y Niebla}, the human rights bulletin published by the \textit{Centro de Investigaci\'on y Educaci\'on Popular/Programa por la Paz}, covering volumes 1 through 36 from 1996 to 2008, available at \url{https://www.nocheyniebla.org}. We also follow \citet{MedinaGallego2011} and \citet{Velez2001} to recover regional information on FARC EP operations. Taken together, these sources provide a more reliable basis for reconstructing the spatial distribution of armed group presence and activity.

Based on the recovered information, we build an incidence matrix for each period in the analysis. Rows correspond to municipalities and columns to armed units. Entries are binary, with a value of 1 indicating civilian targeting by a given armed structure in a municipality during period \textit{t}, and 0 otherwise. From each incidence matrix, we construct a municipal projection network as a weighted undirected graph, where an edge links two municipalities if they were jointly targeted by one or more armed structures during the same period. Edge weights equal the number of distinct armed units that carried out civilian targeting in both municipalities. These matrices form the basis of the network analysis presented in the subsequent sections.

\subsection{Time periods}

This research is organized into periods defined by major political and institutional developments that reshaped the strategies, structures, and territorial presence of armed actors. Instead of imposing evenly spaced intervals, we define periods around critical turning points that altered the dynamics of the conflict. These include shifts in state doctrine and security policy, strategic decisions on territorial expansion by insurgent organizations, and the evolution of paramilitary structures. This division supports a historically grounded approach to modeling changes in the spatial distribution of armed group activity.

The first period begins with the FARC Sixth Conference in 1978, which reaffirmed the group’s military strategy and set the stage for territorial expansion. This phase also coincides with the implementation of the \textit{Estatuto de Seguridad}, associated with increased repression and human rights violations committed by actors linked to state security forces, including \textit{La Mano Negra} and \textit{Los Escuadrones de la Muerte}.

Between 1982 and 1984, the FARC held the Seventh Conference, where it adopted a more explicitly political military strategy, launched its \textit{Plan Estratégico}, and began forming its first mobile columns. At the same time, a fragile truce between insurgent groups and drug trafficking networks began to fracture. Armed confrontation between the FARC and public forces escalated, and the first paramilitary groups, including \textit{Los Escopeteros}, \textit{Los Sanjuaneros}, \textit{Los Tiznados}, and \textit{Las Autodefensas Campesinas del Magdalena Medio}, began to emerge.

The third period, 1985 to 1987, includes the founding of the political party \textit{Unión Patriótica} (UP) and the subsequent wave of selective violence targeting its members. The convergence of armed conflict and narcotrafficking intensified, culminating in the 1985 \textit{Palacio de Justicia} siege by M 19 and the violent state response. These events marked a deepening militarization of the conflict and the erosion of legal political alternatives for the left.

The period from 1988 to 1990 was defined by intensified political violence and a deeper entanglement between armed conflict and organized crime. The assassinations of multiple presidential candidates, together with the systematic targeting of judges, journalists, and local officials, underscored the vulnerability of democratic institutions and the deterioration of public security. During this time, paramilitary activity expanded in rural areas with growing impunity, while guerrilla insurgencies intensified attacks on military and police installations as part of a broader strategy to pressure the State into negotiation. The period closes with the initiation of peace talks involving several guerrilla factions, laying the groundwork for the demobilization processes of the early 1990s.

Between 1991 and 1993, the conflict entered a transitional phase. Several guerrilla and paramilitary groups, such as M 19, the EPL, the PRT, \textit{Quintín Lame}, the ACMM, and \textit{Los Tangueros}, demobilized. The 1990 capture of \textit{Casa Verde}, the FARC main headquarters, signaled a shift in the State’s military posture toward the insurgency. During this time, the FARC strengthened its organizational capacity, while the paramilitary movement began to reorganize after the deaths of key leaders.

The period 1994 to 1996 was marked by the consolidation and expansion of paramilitary forces, often facilitated by elements within the State. The spread of paramilitary violence across several regions coincided with heightened competition over control of drug trafficking routes. These years were characterized by escalating violence and deepening state complicity, which further undermined official efforts to contain the conflict.

From 1997 to 1999, paramilitary groups unified under the banner of \textit{Autodefensas Unidas de Colombia} (AUC), centralizing operations and intensifying counterinsurgency efforts. The AUC territorial expansion coincided with a marked increase in violence against civilians, while peace negotiations with the FARC began, leading to the demilitarization of five municipalities in southern Meta and northern Caquetá.

The 2000 and 2001 period marked the peak of the armed conflict. Violence intensified during ongoing peace talks with the FARC, as the group consolidated a rebel order in territories it had historically held, while paramilitary groups, often operating with the tacit support of public forces, expanded the territorial dispute. In parallel, the national government, backed by US funding, launched a state building strategy known as \textit{Plan Colombia}, focused on military assistance and the dismantling of drug trafficking networks. These dynamics contributed to widespread massacres and forced displacement, severely affecting civilian populations and disrupting local governance.

Between 2002 and 2004, the breakdown of peace negotiations with the FARC, the opening of talks with the AUC, and the implementation of the \textit{Política de Seguridad Democrática} under President Uribe redefined the conflict’s trajectory. Although overall levels of violence began to decline, new intra paramilitary rivalries emerged, particularly in regions where factions split from the AUC. In 2002, the \textit{Bloque Central Bolívar} asserted its autonomy, foreshadowing further fragmentation.

Finally, the 2005 to 2007 period encompasses the formal demobilization of the AUC. Although the process dismantled major paramilitary blocs, it did not end violence. Many former combatants regrouped into emerging criminal organizations, often referred to as \textit{bandas criminales} or BACRIM, which rapidly moved to occupy territories previously held by the AUC.

\subsection{Characterization of bipartite networks}\label{subsec2_2}

This subsection characterizes the bipartite networks over time by examining global graph properties and their relevance to the dynamics of the armed conflict. Visualizations of the complete bipartite networks are reported in the Appendix.

\begin{figure}[!htb]
    \centering
    \subfigure[Armed structures.]  {\includegraphics[scale=0.32]{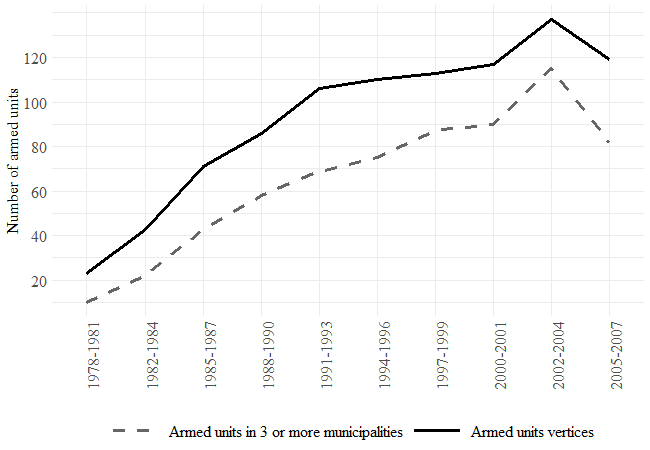}}
    \subfigure[Municipalities.]    {\includegraphics[scale=0.32]{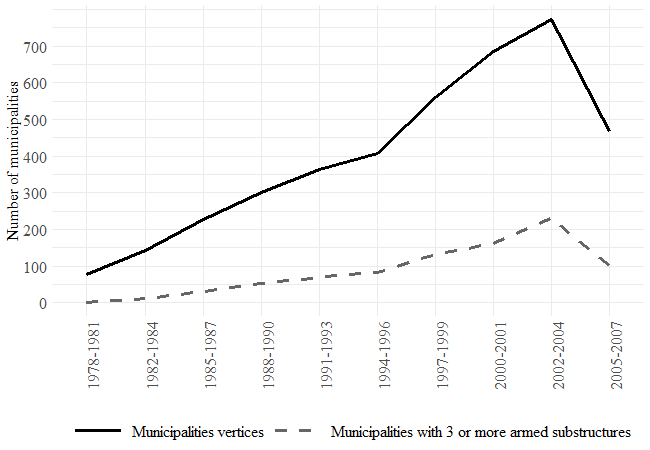}}
    \caption{Order of bipartite networks, 1978-2007.}
    \label{fig:orden}
\end{figure}

\begin{figure}[!htb]
    \centering
    \includegraphics[scale=0.5]{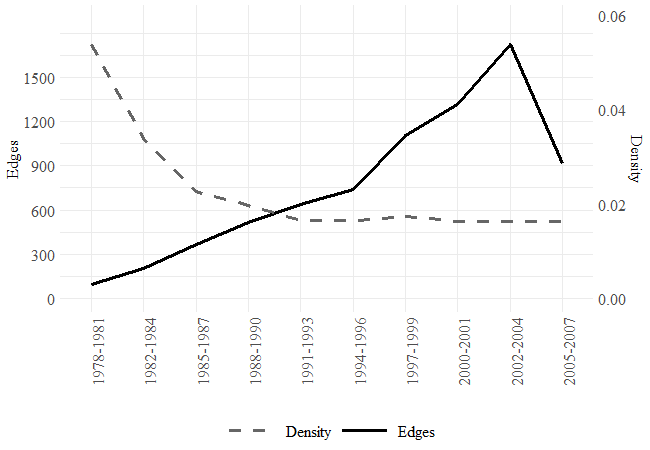}
    \caption{Size and density of bipartite networks, 1978-2007.}
    \label{fig:Tamaño}
\end{figure}

Figures \ref{fig:orden} and \ref{fig:Tamaño} show the evolution of order, size, and density of the bipartite networks across periods. Overall, we observe a sustained increase in the number of nodes and edges, reflecting the expansion of armed group activity in both organizational and territorial terms. This growth is particularly pronounced during the first decade of the study. The number of FARC units increased nearly fourfold, rising from 14 fronts in 1978 to 1981 to 51 by the end of 1988 to 1990. This expansion is consistent with the strategic transformations initiated at the Seventh Conference in 1982. Paramilitary growth followed a parallel but slightly delayed trajectory, increasing from 8 documented units in the initial period to 30 by 1988 to 1990, driven largely by the proliferation of regionally anchored self defense groups and counterinsurgency projects. Although some right wing structures demobilized between 1991 and 1993, the number of their units continued to grow until 1996, after which a national paramilitary project consolidated. Over the same years, the FARC also expanded, and by the end of the decade its structure included 35\% more fronts and mobile columns.

The number of armed units reached its peak in 2002 to 2004, which also corresponded to the FARC’s largest organizational development. At that point, the group, particularly its \textit{Bloque Oriental}, recorded the highest number of combatants in its history. On the paramilitary side, the increase in structures reflected internal splits that emerged during the peace negotiations promoted by President Uribe’s administration. The \textit{Bloque Central Bolívar}, together with substructures loyal to its commanders, declared independence and formed a separate organization with its own chain of command, contributing to the proliferation of paramilitary units. We then observe a substantial decline in 2005 to 2007 as several AUC substructures demobilized. In parallel, the FARC faced a major setback after the launch of \textit{Plan Patriota}, a large-scale military operation led by the Colombian Army. The campaign dismantled eleven guerrilla fronts surrounding the capital and contributed to the overall reduction in armed units. Although the FARC remained capable of sustaining the war effort, its strategy of besieging the national government was no longer attainable.

Similarly, the number of municipalities affected by victimization increased steadily until 2002 to 2004, after which the territorial presence of both actors declined, particularly in the central region. Network density, which remained low throughout, decreased until 1997 and then stabilized. Despite growth in nodes and edges, the network became more dispersed, consistent with interactions spreading across a broader set of municipalities and substructures. Overlap across locations increased, but municipalities with three or more substructures grew more slowly than those with one or two. Thus, while armed presence expanded nationally, intense overlap widened more gradually, suggesting that the most complex territorial disputes were concentrated in specific regions even as the set of affected municipalities continued to grow.

\begin{figure}[!htb]
    \centering
    \subfigure[Degree, armed structures.]  {\includegraphics[scale=0.32]{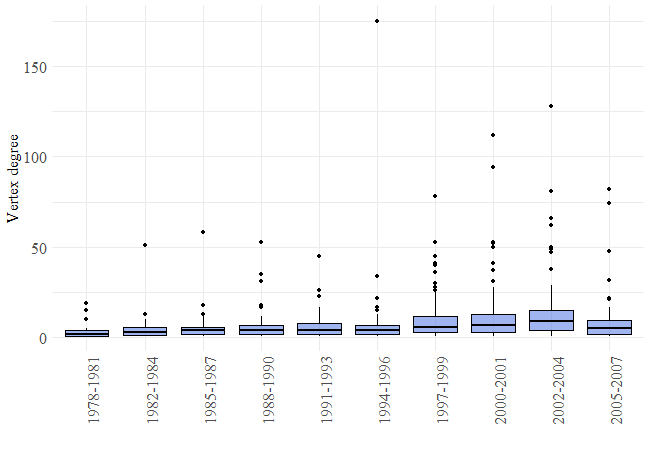}}
    \subfigure[MDegree, municipalities.]   {\includegraphics[scale=0.32]{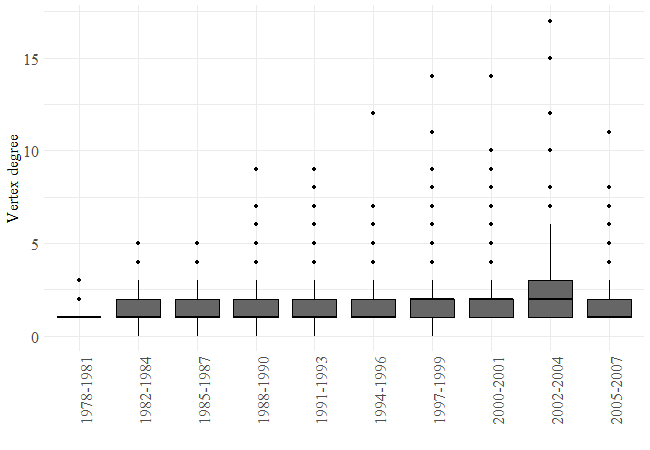}}
    \caption{Degree in bipartite networks, 1978--2007.}
    \label{fig:FuerzaGradoRB}
\end{figure}

Figure \ref{fig:FuerzaGradoRB} shows the degree distribution for armed structures and municipalities in the bipartite systems. The average number of municipalities in which a given armed group operated increased from 1978 to 1993, with the mean rising from 4.09 to 6.03 and the third quartile from 4 to 8. Starting in 1994, with the formation of \textit{Las Autodefensas Campesinas de Córdoba y Urabá}, we observe a clear structural shift. In this period, the group expanded to 175 municipalities, far above the upper range of most other organizations, 75\% of which operated in at most 7 municipalities. This contrast reflects the territorial reach of the ACCU relative to other armed actors. From that point onward, degree asymmetry becomes more pronounced, with increasingly extreme outliers. This is especially evident in 2002 to 2004, when \textit{El Bloque Norte} of the AUC emerges as the most expansive substructure. In subsequent periods, the number of highly expansive substructures declines, consistent with increasing fragmentation during and after demobilization.

A similar pattern emerges for the municipal side of the bipartite network. The average vertex degree increased from 1978 to 2004, rising from 1.24 to 2.23, while the maximum increased from 3 to 17. This growing asymmetry suggests increasing centralization, where a small number of highly connected municipalities acted as territorial hubs. In the early periods, the municipalities concentrating most connections were small cities clustered mainly in the \textit{Magdalena Medio} region. Between 2002 and 2004, the highest degree vertices correspond to the country’s largest cities, consistent with the expansion of the conflict from rural areas to urban centers.

\begin{figure}[!htb]
    \centering
        \includegraphics[scale=0.5]{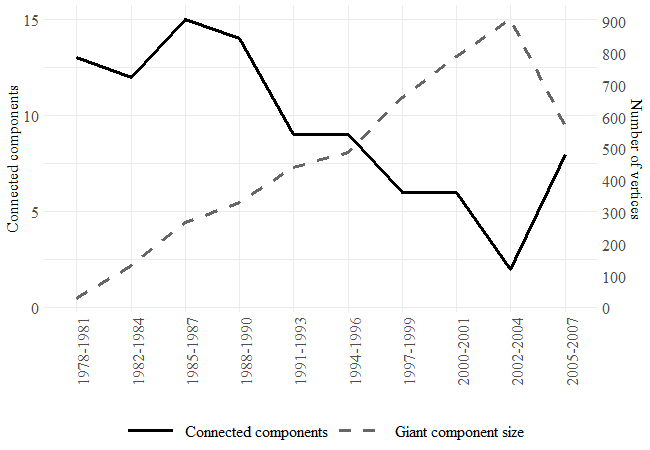}
        \caption{Number of components and size of the giant component, 1978-2007.}
        \label{fig:Componentes}
\end{figure}

To assess territorial overlap among armed groups, it is useful to examine changes in the number of connected components in the bipartite networks. A larger component count indicates more spatially isolated operations, whereas a smaller count reflects greater interconnection. Figure \ref{fig:Componentes} reports the number of components and the size of the giant component across periods. In the first period, most armed groups, particularly FARC units, occupied small disconnected clusters. By the second period, some components began to centralize around paramilitary structures linked to multiple municipalities. This was followed by expansion driven by the FARC Seventh Conference, which promoted the creation of new fronts by splitting existing ones, extending presence into additional territories with limited competition. As the conflict intensified, the number of separate clusters declined steadily and the giant component grew continuously, reaching by 2004 a pattern consistent with increasing structural integration. This trend reverses in the final period, when paramilitary groups undergo substantial fragmentation.

\subsection{Characterization of the municipality projection}\label{subsec2_4}

\begin{figure}[!htb]
\centering
    \includegraphics[width=0.75\linewidth]{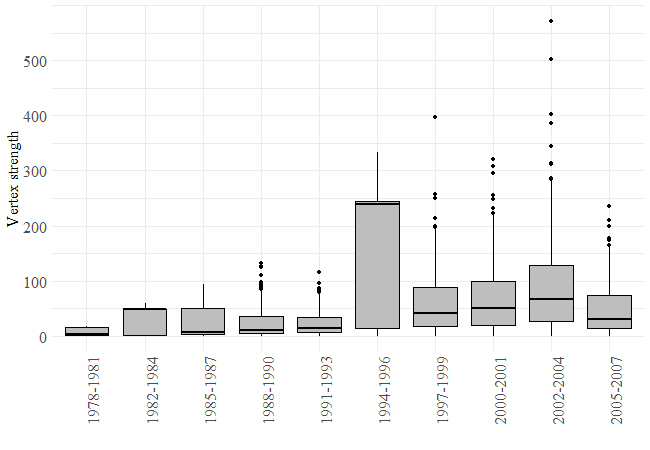}
    \caption{Strength of vertices in the municipality projections, 1978-2007.}
    \label{fig:FuerzaGradoActores}
\end{figure}

To analyze the territorial configurations induced by interactions among armed actors, we examine the municipal projections of the bipartite system, following the approach of the previous section. Figure \ref{fig:FuerzaGradoActores} shows the strength distributions for the municipal networks. From a low connectivity system in 1978, where vertices are only weakly linked, the topology shifts toward a left skewed distribution. From 1985 onward, we observe a marked increase in extreme outliers and a rise in the median, which doubles between the third and fifth periods. This pattern later reverses as a subset of actors becomes dominant, generating dense and overlapping subgraphs in the municipal projection. These high order cohesive groupings follow directly from the projection, since the broader the territorial scope of an actor in the bipartite graph, the larger the resulting cluster of interconnected municipalities in the one mode projection. Thus, during periods of high territorial mobility, network strength tends to increase, reflecting stronger structural integration across conflict affected regions. This phase of expansion is followed by a pronounced shift between 1997 and 2007, as the distribution becomes increasingly right skewed. Although the median strength continues to rise, upper quartiles and outliers grow faster, and a small number of municipalities concentrate a substantial share of ties while many others remain sparsely connected.

We assess the plausibility of power law behavior in the strength distributions, of the form \(p(x) \propto x^{-\alpha}\), using the methodology introduced by \citet{clauset2009} and implemented in the \texttt{poweRlaw} package \citep{gillespie2015}. A goodness of fit test yields a \(p\) value by comparing the distance between the empirical distribution and the fitted power law with the same distance computed on synthetic datasets generated from the fitted model. The \(p\) value is the proportion of synthetic Kolmogorov Smirnov distances that exceed the observed distance between the model and the data. Large values suggest that the power law provides an adequate fit, while small values indicate poor fit. \citet{clauset2009} recommend rejecting the power law hypothesis when \(p \leq 0.1\), meaning that the observed discrepancy would be unlikely to arise from random variation under the fitted model.

For each period, we estimate the lower bound \(x_{\min}\) that defines the tail region and compute the scaling exponent \(\hat{\alpha}\) by maximum likelihood. To evaluate fit above \(x_{\min}\), we use a semi parametric bootstrap. We generate 5,000 synthetic datasets that preserve the empirical distribution below \(x_{\min}\) and draw observations above \(x_{\min}\) from the fitted power law. Each synthetic dataset is obtained by resampling values below the threshold from the observed data and sampling the remainder from the fitted model. For each sample, we re estimate \(x_{\min}\) and \(\hat{\alpha}\) and compute the corresponding Kolmogorov Smirnov distance. The proportion of synthetic distances that exceed the empirical distance is the resulting \(p\) value.

\begin{table}[!htb]
\centering
\caption{Power-law fit parameters for strength distributions.}
\label{tab:powerlaw_parameters}
\begin{threeparttable}
\small
\begin{tabular*}{\columnwidth}{@{\extracolsep\fill}lcccccc@{\extracolsep\fill}}
\toprule
Period & $n$ & $n_{\text{tail}}$ & $\hat{\alpha}$ & $x_{\min}$ & KS stat. & $p$-value \\
\midrule
1985--1987 & 208 & 58  & 7.390 & 57  & 0.11637 & 0.035 \\
1988--1990 & 259 & 16  & 6.596 & 89  & 0.10692 & 0.754 \\
1991--1993 & 345 & 47  & 5.465 & 48  & 0.06802 & 0.608 \\
1997--1999 & 553 & 31  & 6.737 & 116 & 0.10407 & 0.290 \\
2000--2001 & 680 & 102 & 4.441 & 112 & 0.08677 & 0.027 \\
2002--2004 & 771 & 140 & 6.107 & 129 & 0.05906 & 0.149 \\
2005--2007 & 458 & 102 & 3.836 & 76  & 0.10708 & 0.001 \\
\bottomrule
\end{tabular*}
\begin{tablenotes}
\item \textit{Note:} $\hat{\alpha}$ is the scaling exponent estimated via maximum likelihood above $x_{\min}$; $n_{\text{tail}}$ indicates the number of observations in the tail.
\end{tablenotes}
\end{threeparttable}
\end{table}

As shown in Table \ref{tab:powerlaw_parameters}, the power law hypothesis cannot be rejected in four of the seven periods, 1988 to 1990, 1991 to 1993, 1997 to 1999, and 2002 to 2004, where \(p\) values exceed the 0.1 threshold. In these cases, the observed deviations from the model are not statistically significant. Caution is warranted, since between 1988 and 1993 the number of observations in the tail is small, which limits test power and reduces the ability to detect departures from a power law \citep{clauset2009}. Under these conditions, failure to reject does not confirm power law behavior, it may reflect limited information. By contrast, 2000 to 2001 and 2005 to 2007 yield low \(p\) values, providing stronger evidence against power law scaling in those years.

To assess whether the power law fits better than plausible alternatives, we use Vuong’s likelihood ratio test to compare it with log normal, exponential, and Poisson models (Table`\ref{tab:powerlaw_tests}).\footnote{Vuong’s test relies on the normalized log likelihood ratio between two competing models. The statistic is computed as the average of the pointwise log likelihood ratios divided by their standard deviation, \(\text{LR} = \frac{1}{n} \sum_{i=1}^{n} \log\!\left(\frac{f_1(x_i)}{f_2(x_i)}\right) \Big/ \text{SD}\), where \(f_1\) and \(f_2\) denote the fitted densities and \(\text{SD}\) is the standard deviation of the log likelihood differences.} The test yields a \(p\) value assessing whether the observed sign of the statistic reflects a meaningful difference in fit or is compatible with random variation. When this \(p\) value is small, for example \(p < 0.1\), the sign is considered informative and indicates which model fits better. When the \(p\) value is large, the evidence is insufficient to favor one model.

The Vuong tests indicate that a Poisson model is strongly disfavored in every period as a single–parameter description of the marginal strength distribution, which is consistent with the pronounced heavy–tailed heterogeneity observed across municipalities. A small number of municipalities emerge as hubs in which multiple armed structures overlap, plausibly reflecting persistent asymmetries in territorial control and dispute intensity together with path-dependent connectivity, rather than ties forming at random. Importantly, this evidence against a marginal Poisson for node strengths does not preclude the use of a Poisson SBM for the edge weights, since conditional on block structure the SBM induces heterogeneous interaction rates across community pairs and can accommodate substantial variability in expected tie intensities, while the log-normal and exponential specifications remain competitive alternatives for the marginal strength distribution.
 
The Poisson model is strongly disfavored as a one parameter model for the marginal strength distribution in every period, consistent with heavy tailed heterogeneity across municipalities; this does not rule out Poisson modeling for edge weights conditional on block structure, since an SBM allows heterogeneous interaction rates across community pairs.

\begin{table}[!htb]
\centering
\caption{Likelihood ratio tests comparing the power-law fit to alternative distributions.}
\label{tab:powerlaw_tests}
\begin{threeparttable}
\small
\begin{tabular*}{\columnwidth}{@{\extracolsep\fill}lcccccc@{\extracolsep\fill}}
\toprule
Period & \multicolumn{2}{c}{Log-normal} & \multicolumn{2}{c}{Exponential} & \multicolumn{2}{c}{Poisson} \\
\cmidrule(lr){2-3} \cmidrule(lr){4-5} \cmidrule(lr){6-7}
& LR & $p$ & LR & $p$ & LR & $p$ \\
\midrule
1988--1990 & -0.227 & 0.590 & -0.102 & 0.541 & 1.987 & 0.023 \\
1991--1993 & -0.448 & 0.673 & -0.091 & 0.536 & 1.655 & 0.049 \\
1997--1999 & -0.660 & 0.745 & -0.999 & 0.841 & 1.932 & 0.027 \\
2002--2004 & -0.889 & 0.813 & -0.620 & 0.732 & 4.779 & 0.000 \\
\bottomrule
\end{tabular*}
\begin{tablenotes}
\item \textit{Note:} LR denotes the normalized log-likelihood ratio statistic from Vuong’s test comparing the power-law model to the alternative. Positive values favor the power-law; negative values favor the alternative. The $p$-values indicate whether these differences are statistically significant.
\end{tablenotes}
\end{threeparttable}
\end{table}

\begin{figure}[!htb]
    \centering
    \includegraphics[width=0.65\linewidth]{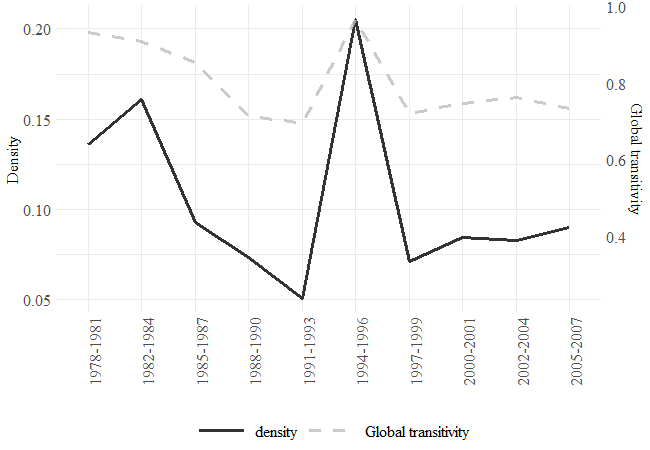}
    \caption{Density and transitivity in the municipality networks, 1978-2007.}
    \label{fig:Den_Tr_mun}
\end{figure}

When examining density and transitivity in the municipal networks, we observe a marked gap between them. Transitivity ranges from moderate to high, between 0.3 and 0.95, while density remains below 0.2 across the ten periods, see Figure \ref{fig:Den_Tr_mun}. This pattern suggests a structure dominated by cohesive groups that are weakly connected to one another, except in 1994 and 1996. As noted above, this interval corresponds to the expansion of a small subset of armed structures, whose induced municipal cliques subsume those associated with the zones of influence of other actors. Both indicators reach their maximum during this period.

\begin{figure}[!htb]
    \centering
        \includegraphics[width=0.65\linewidth]{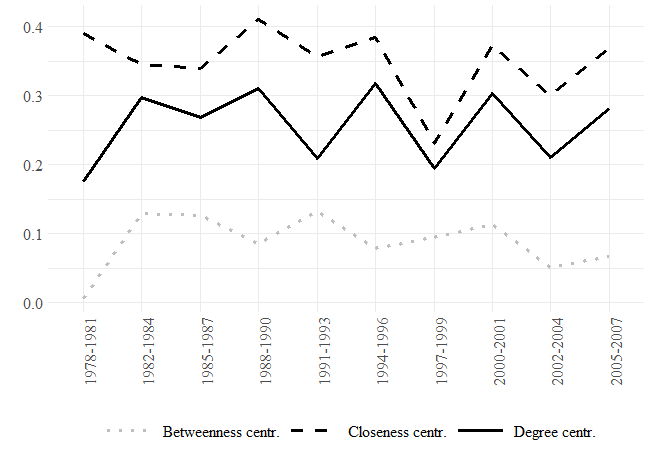}
        \caption{Centralization indices in the municipality networks, 1978-2007.}
        \label{fig:CentrMpios2}
\end{figure}

As shown in Figure \ref{fig:CentrMpios2}, we observe moderate to high closeness centralization together with low betweenness centralization. This pattern indicates a structure in which a few municipalities are at short topological distance from many others, yet they do not function as essential bridges. It arises when large and overlapping zones of influence, each inducing cliques in the projection, converge in specific municipalities and position them as hubs. These are locations where expansive actors coincide. Such municipalities are adjacent to all others within each actor’s zone and therefore to a large subset of the network. At the same time, because actor induced cliques are densely connected, the network offers many alternative paths between most pairs of nodes as overlap increases. The persistently low betweenness centralization across periods suggests that territorial contestation does not hinge on a small set of strategic municipalities. Instead, concurrence among actors is more broadly distributed across space, pointing to a diffuse pattern of dispute.

\begin{figure}[!htb]
    \centering
    \subfigure[Clique number and number of maximal cliques.]  {\includegraphics[scale=0.32]{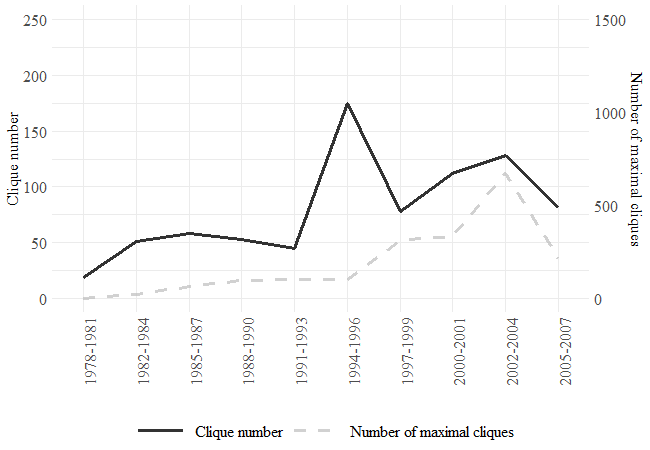}}
    \subfigure[Distribution of maximal clique size.]   {\includegraphics[scale=0.32]{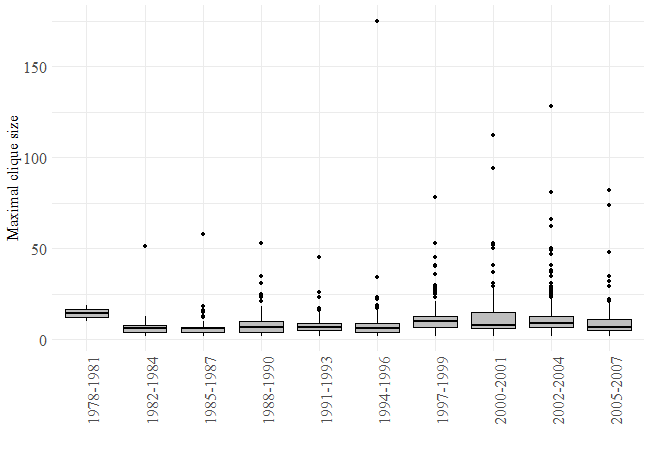}}
    \caption{Cliques in the municipality networks, 1978-2007.}
    \label{fig:CliqueSize}
\end{figure}

Figure \ref{fig:CliqueSize} shows the number of components and the size of the giant component in the municipal networks. During the first decade of the study, the large number of components indicates that actors operated in largely separate subgraphs. From 1991 onward, isolated components decline rapidly and the giant component expands, reinforcing the view that territorial convergence increased over time.

We also observe changes in clique size across periods, which reveal a clear pattern of territorial expansion by armed structures (Table \ref{tab_cliques}). In 1978 to 1984, cliques were relatively small, with an average size of 4.303 municipalities and a maximum of 49, suggesting a limited and localized presence. Over time, especially from the late 1980s onward, both the median and the maximum clique size increase. This growth becomes particularly pronounced in 1994 to 1996, when the average clique size reaches 11.17 municipalities and the maximum rises to 241. As noted above, this aligns with the expansion of a dominant structure.

\newpage 

The long tail of the maximal clique size distribution indicates that large cliques are not rare. This implies that several armed structures have extensive zones of influence, which is consistent with a broader set of influential actors and lower concentration of territorial reach in a single group. These outliers become more common in later periods, notably 1997--1999 and 2002--2004, in line with intensified territorial disputes. This pattern suggests that more structures have the capacity to operate across a larger number of locations. By the end of the study period, in 2005--2007, the average clique size decreases slightly to 10.71, yet the maximum remains high at 105 municipalities. This indicates that, even as overall expansion declined with demobilization, structures capable of sustained multi municipal operations persisted.

\begin{table}[!htb]
\centering
\caption{Proportion of municipalities in multiple cliques.}\label{tab_cliques}
\label{tab:clique_overlap}
\begin{threeparttable}
\small
\begin{tabular}{lccc}
\toprule
Period & $>1$ clique & $>10$ cliques & $>30$ cliques \\
\midrule
1978--1981 & 0.21 & 0.00 & 0.00 \\
1982--1984 & 0.33 & 0.00 & 0.00 \\
1985--1987 & 0.40 & 0.02 & 0.00 \\
1988--1990 & 0.44 & 0.08 & 0.00 \\
1991--1993 & 0.41 & 0.01 & 0.00 \\
1994--1996 & 0.42 & 0.04 & 0.00 \\
1997--1999 & 0.51 & 0.16 & 0.05 \\
2000--2001 & 0.50 & 0.16 & 0.03 \\
2002--2004 & 0.56 & 0.26 & 0.08 \\
2005--2007 & 0.46 & 0.13 & 0.00 \\
\bottomrule
\end{tabular}
\begin{tablenotes}
\item \textit{Note:} Each cell reports the proportion of municipalities belonging to more than one, ten, or thirty cliques per period.
\end{tablenotes}
\end{threeparttable}
\end{table}

In earlier periods, when cliques were relatively small and the share of municipalities belonging to more than one clique was low, the conflict displayed a segmented territorial structure in which armed actors operated in discrete and largely non overlapping zones of influence. This pattern reflects spatial partitioning, with territorially exclusive control and municipalities rarely claimed by more than one group. Over time, this configuration shifted toward a more fragmented landscape. From the mid 1990s onward, and more sharply in the 2000s, the proportion of municipalities belonging to multiple cliques increased steadily, reaching 56\% in 2002 to 2004. This reflects not only territorial expansion but also a substantial intensification of overlap among rival actors. By that point, more than a quarter of municipalities appeared in more than ten cliques, and nearly one in ten appeared in more than thirty. These dense intersections indicate a territorial order characterized by competition, instability, and contested sovereignty, where multiple armed structures coexisted within the same municipalities. Rather than dominance by a single group, the evidence points to a multipolar configuration of violence with fluid and overlapping claims to space.

\section{Modeling the projection networks and detecting communities}

Projections of bipartite networks often exhibit structures formed by multiple complete subgraphs, some overlapping and others disjoint. This pattern arises because nodes that share neighbors in the bipartite graph become fully connected in the projection. In conflict networks, the projection highlights zones with different intensities of territorial dispute. Municipalities linked to the same armed unit form complete subgraphs that represent areas of influence. When different armed groups operate in the same municipalities, the corresponding subgraphs overlap, creating regions of higher connectivity. These shared influence areas are denser and involve stronger ties, which increases the probability that municipalities cluster into the same community. By contrast, municipalities where a single armed structure targeted civilians tend to form separate subgraphs, sometimes isolated, which delineates regions where competition is limited or expressed mainly through direct military operations.

These projected structures inform the dynamics of territorial competition and strategic interaction among armed actors. When overlap is pervasive, the network suggests stronger dispute and shifting alignments. When disjoint complete subgraphs predominate, the projection reflects a more segmented territorial arrangement. The balance between overlap and separation captures the fragmentation and fluidity of territorial influence.

The descriptive analysis reveals a complex and evolving topology in which overlapping and disjoint subgraphs reflect different degrees of territorial dispute. To assess how these features shape municipal clustering, we fit a set of stochastic block models with \textsf{Poisson} edges, which characterize community structure in weighted networks by modeling interaction intensities between nodes. This approach evaluates whether municipalities under armed influence form distinct communities or remain strongly interconnected, providing a systematic assessment of territorial fragmentation and consolidation. It also provides model-based uncertainty summaries through the estimated membership probabilities (responsibilities) produced by the variational fit. Accordingly, we interpret these quantities as soft clustering outputs from variational estimation, rather than as draws from an MCMC posterior.

Using the function \texttt{BM\_poisson} from the \texttt{blockmodels} package \citep{Borgatti2016} in \texttt{R} \citep{R2019}, we fit, for each period, multiple \textsf{SBMP} specifications to the municipal networks under different numbers of communities \(Q\). We select the optimal partition using the integrated classification likelihood (ICL) criterion \citep{Biernacki2000}, which balances fit and parsimony and is tailored to clustering problems \citep{KolaczykCsardi2014}. The selected values of \(Q\) and the corresponding community compositions are shown in Figure \ref{fig:sankey_municipalities}. Additional results are reported in Appendix \ref{app:icl}.

\begin{figure}[!htb]
\centering
\includegraphics[angle=270, width=0.72\textwidth]{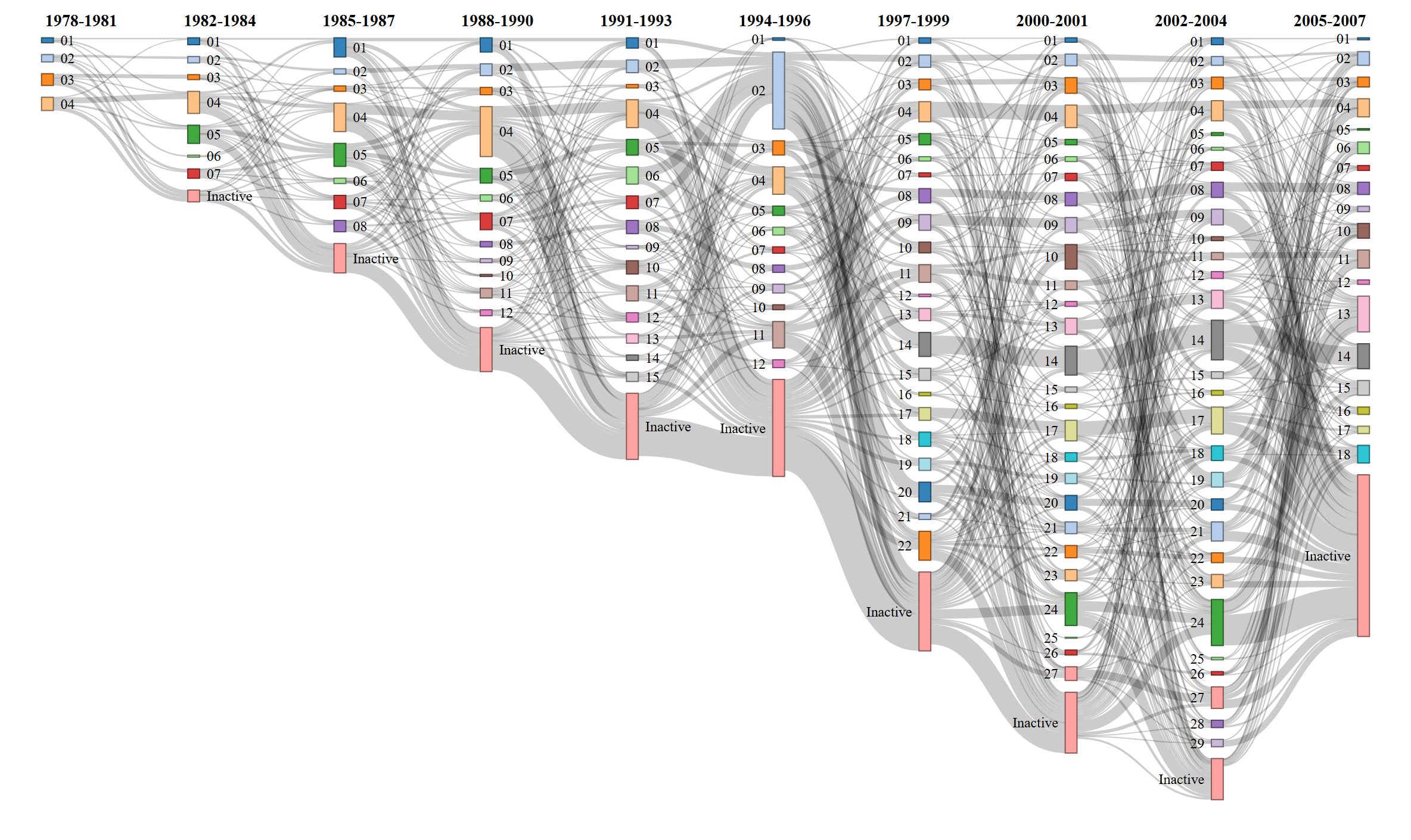}
\caption{Communities in the municipality network, 1982–2007.}
\label{fig:sankey_municipalities}
\end{figure}

The diagram tracks the evolution of the communities over time, highlighting continuities and splits. Each vertical column corresponds to a period and displays the detected communities. The flows between columns represent transitions of municipalities across clusters. Flow width is proportional to the number of municipalities that move between communities, which highlights continuity and realignment. Some municipalities appear as ``inactive'' in certain periods, indicating that they were previously present but later dropped from the network, either temporarily or permanently. This visualization summarizes how municipal clustering evolves and how territorial interactions change over the course of the conflict.

The number of communities increases over time, except in 1994 to 1996 when the network topology shifts. The structure evolves from four communities in the first period to 33 in 2002 to 2004. During the first decade, the number of detected communities increases about fourfold. The largest increase occurs in the late 1990s, when \(Q\) rises from 12 in 1994 to 1996 to 22 in 1997 to 1999, and continues to grow to its peak in 2002 to 2004. By 2005 to 2007, \(Q\) declines to 18, consistent with the demobilization of paramilitary groups.

Throughout the study period, the distribution of municipalities across communities changed substantially. In the first period, the distribution was relatively even, with two communities containing about 35\% of the vertices each, a third about 20\%, and the fourth about 10\%. By contrast, during the 1980s a single dominant community contained a large share of municipalities. In later years, the size distribution became more balanced, dominance declined, and municipalities were spread across a larger number of smaller communities.

\begin{figure}[!htb]
    \centering
    \subfigure[Community interaction intensities, $\lambda_{ij}$.]  {\includegraphics[scale=0.4]{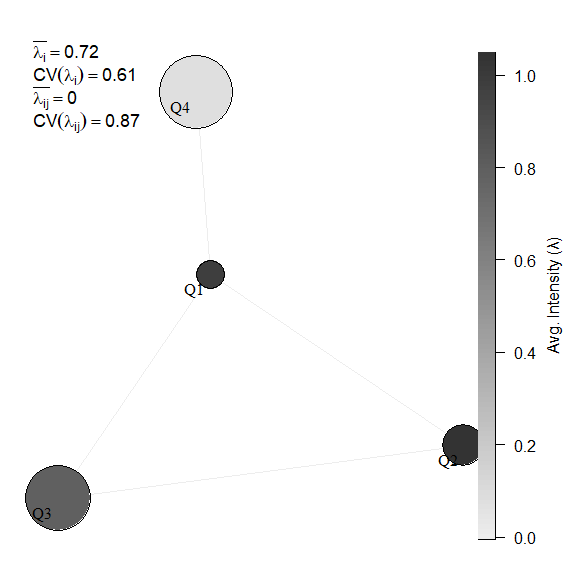}}
    \subfigure[Municipality network.]   {\includegraphics[scale=0.4]{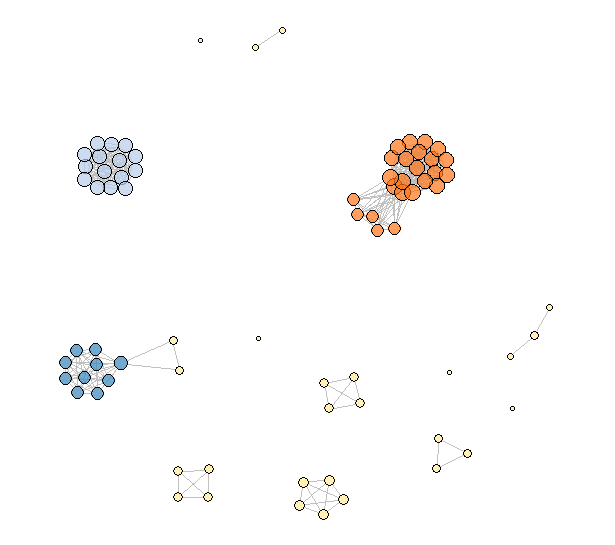}}
    \caption{Communities in the municipality network, 1982-1984.}
    \label{fig:cluster_interactions_I}
\end{figure}

Between 1978 and 1981, the network consisted of thirteen disconnected components. The SBM grouped them into four communities with a sharp within between contrast, consistent with localized armed groups operating in bounded and non overlapping zones. The average within community interaction intensity was below 1, with \(\bar{\lambda}_{i} = 0.72\), while between community interactions were essentially zero, with \(\bar{\lambda}_{ij} = 0\). Community assignments are shown in Figure \ref{fig:cluster_interactions_I}. Colors indicate membership, edge styles distinguish within community, shown as solid, from between community, shown as dotted, ties, and node sizes are proportional to the log transformed number of victims.

\begin{figure}[!htb] 
    \centering
    \includegraphics[scale=0.75]{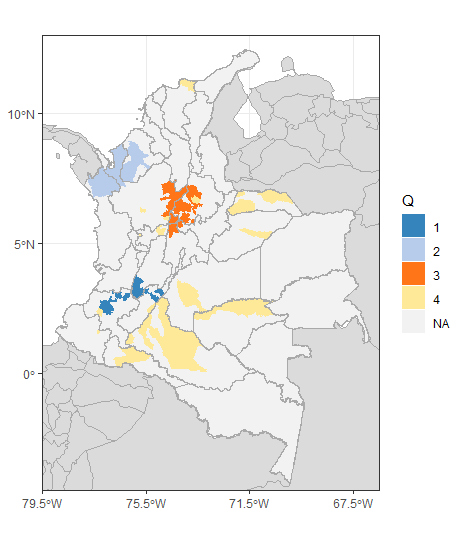} 
    \caption{Communities in the municipality network, 1982-2007.}
    \label{municipalitiesmap_communities_1978-1981}
\end{figure}

Within this arrangement, we identify three main configurations. First, \(Q_2\) is a compact enclave in \textit{Urabá}. Within class ties are uniform and close to the single overlap benchmark, consistent with a zone organized around the V Front, where banana sector unionism and labor mobilization provided dense social infrastructure, while \textit{Defensa Civil} units were legalized in parallel. The result is strong internal cohesion with limited external contact. Second, \(Q_3\) forms a contested corridor in the \textit{Magdalena Medio}. Ties are dense but markedly heterogeneous across municipalities, reflecting overlap between the IV and XI Fronts and early paramilitary formations, including \textit{Los Sanjuaneros}, \textit{Autodefensas de Yacopí}, and \textit{Mano Negra}. This mix produces short high weight subgraphs with only sparse bridges outward. Third, \(Q_1\) corresponds to an intermediate stronghold along the southern \textit{Tolima} and northern \textit{Cauca} and \textit{Huila} belt. Density remains high and internal variation is moderate, consistent with corridors where the VI and VIII Fronts expanded through peasant organizations, yielding stable within class connectivity with limited reach beyond adjacent municipalities. The resulting spatial distribution is shown in Figure \ref{municipalitiesmap_communities_1978-1981}.

Outside these clusters, the peripheral structure \(Q_4\) consists of small disconnected cliques and dyads across the \textit{Meta}, \textit{Caquetá}, and southern \textit{Cauca} colonization frontiers. In this setting, State absence and recent settlement facilitated local regulatory roles by FARC fronts, producing geographically isolated enclaves that appear as small internally complete subgraphs with minimal cross class exchange. This pattern reflects fragmented strongholds rather than a contiguous area of control.

\begin{figure}[!htb]
    \centering
    \subfigure[Community interaction intensities, $\lambda_{ij}$.]{
        \includegraphics[scale=0.4]{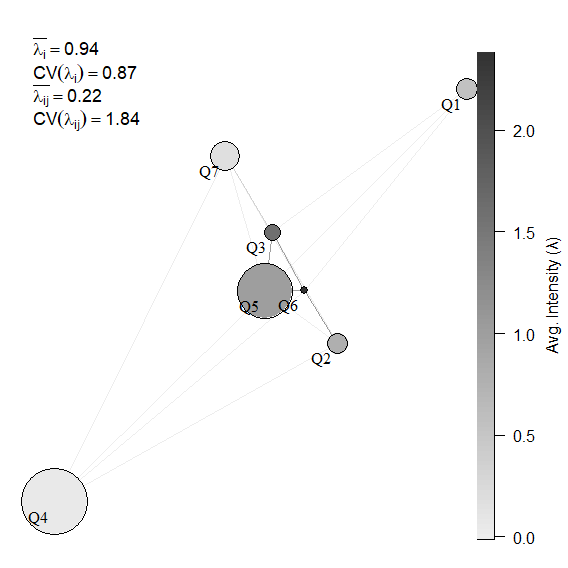}
    }
    \subfigure[Geographical distribution of communities.]{
        \includegraphics[scale=0.4]{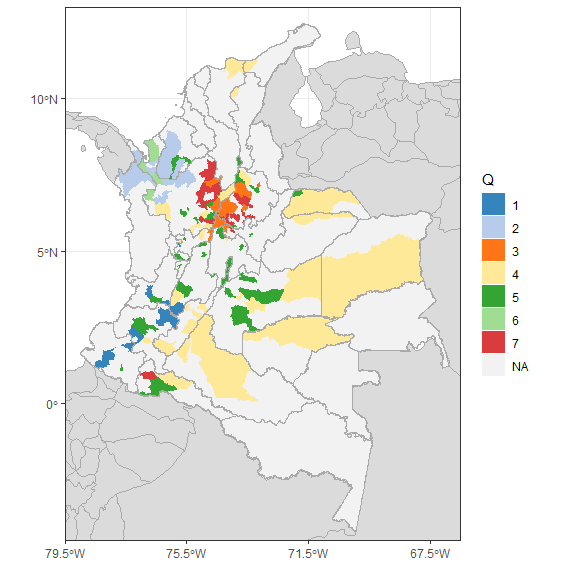}
    }
    \subfigure[Municipality network.]{
        \includegraphics[scale=0.4]{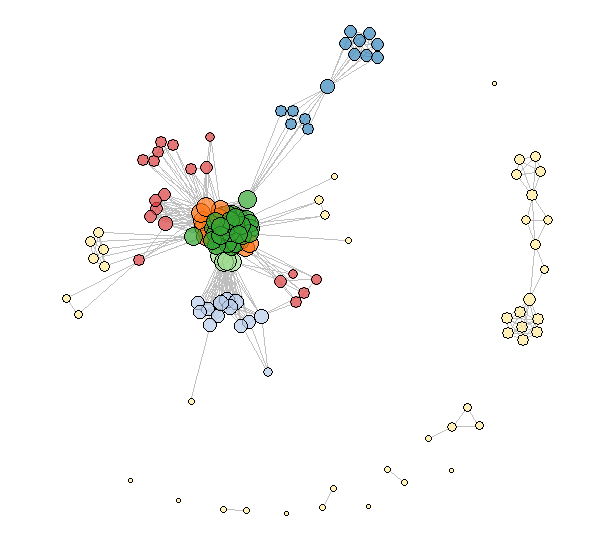}
    }
    \subfigure[Network core.]{
        \includegraphics[scale=0.4]{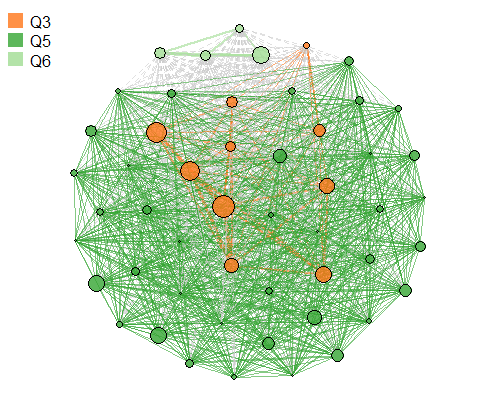}
    }
    \caption{Communities in the municipality network, 1982--1984.}
    \label{fig:cluster_interactions_Ia}
\end{figure}

Between 1982 and 1984, the municipal projection network shifts from scattered classes to a structure dominated by a giant component. This change coincides with two parallel dynamics, the FARC Seventh Conference, which set the strategy of transforming the organization into a national army to seize state power, and the strengthening of paramilitary efforts as drug traffickers, landowners, local elites, and sectors of the security forces consolidated coordinated networks. As armed units expanded into additional municipalities, overlap in their operating footprints connected previously separate clusters into a single component, leaving only a small peripheral set outside it. Figure \ref{fig:cluster_interactions_Ia} summarizes the partition. Nodes represent communities scaled by their municipal share, darker shading indicates higher average within-community intensity, and edge colors encode mean between community intensities.

Against this background, the SBM yields seven communities organized around a central triad, \(Q_5\), \(Q_6\), and \(Q_3\), where between community interaction strengths \(\bar{\lambda}_{qr}\) are relatively high and municipal weight is concentrated. This configuration is driven by the broad footprint of MAS. When MAS overlaps with distinct secondary actors, the algorithm separates the resulting subgraphs into multiple communities connected through MAS. Community \(Q_5\) aggregates MAS municipalities across non contiguous regions, with moderate internal cohesion and stronger ties to the other core sets than to the periphery, consistent with multi regional reach. Community \(Q_6\), centered in \textit{Urabá}, combines MAS with the FARC Fifth Front, \textit{Defensa Civil}, and \textit{Mano Negra}. Despite its small size, it shows the strongest internal coupling and the tightest attachment to the core. Community \(Q_3\), located in \textit{Magdalena Medio}, is less cohesive internally but retains above peripheral between community intensities with \(Q_5\) and \(Q_6\), anchoring its role in the triad.

Outside the core, \(Q_1\) forms a compact southwestern cluster under Fronts VI and VIII with limited outward links. \(Q_2\) in \textit{Urabá} and \(Q_7\) in \textit{Magdalena Medio} show intermediate internal intensity and weak connections to the core, grouping municipalities where actors that co-occur with MAS in \(Q_5\) operate in its absence. \(Q_4\) collects the widest periphery, with numerous low intensity single unit overlaps across dispersed localities, consistent with front expansion under the Seventh Conference line that remained locally bounded and only loosely integrated into system wide exchange.

As a result of peace negotiations, the \textit{Acuerdo de La Uribe}, a ceasefire, was signed in 1984. A year later, the \textit{Unión Patriótica} (UP) was created to provide a legal platform for historically excluded left wing movements. The truce quickly unraveled. Lacking firm political and institutional backing, it was followed by a paramilitary campaign targeting UP militants, unionists, social leaders, and peasant organizers. Neither side adhered to the ceasefire. In this context, the conflict expanded on both fronts. FARC, under centralized command, roughly doubled its number of fronts, while paramilitary formations, financed by local elites and drug traffickers, spread across the country.

\begin{figure}[!htb]
    \centering
    \subfigure[Community interaction intensities, $\lambda_{ij}$.]{
        \includegraphics[scale=0.4]{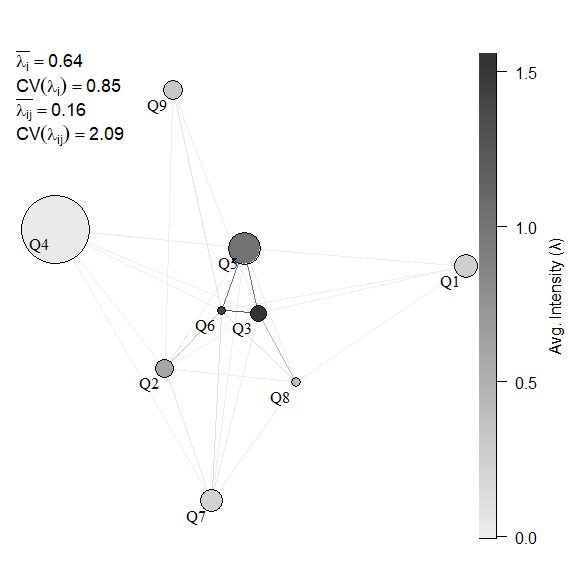}
    }
    \subfigure[Geographical distribution of communities.]{
        \includegraphics[scale=0.4]{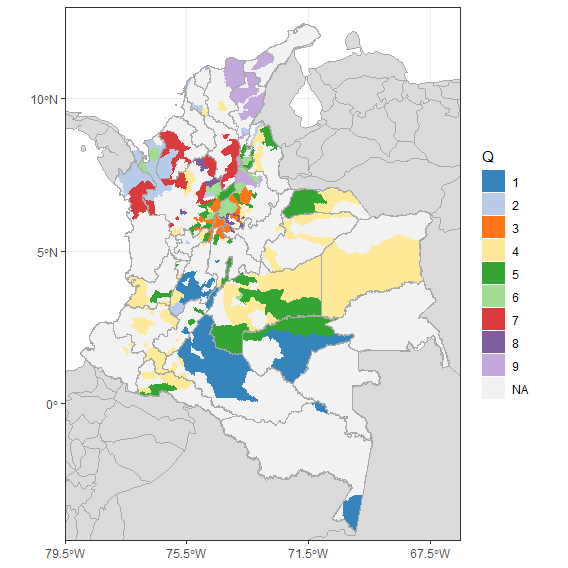}
    }
    \subfigure[Municipality network.]{
        \includegraphics[scale=0.4]{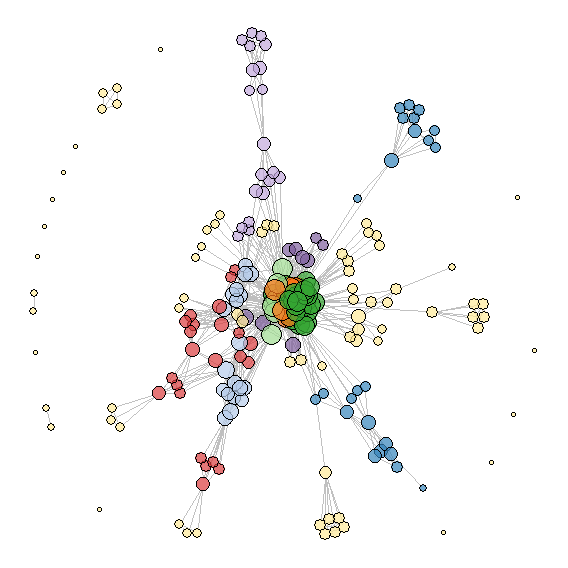}
    }
    \subfigure[Network core.]{
        \includegraphics[scale=0.4]{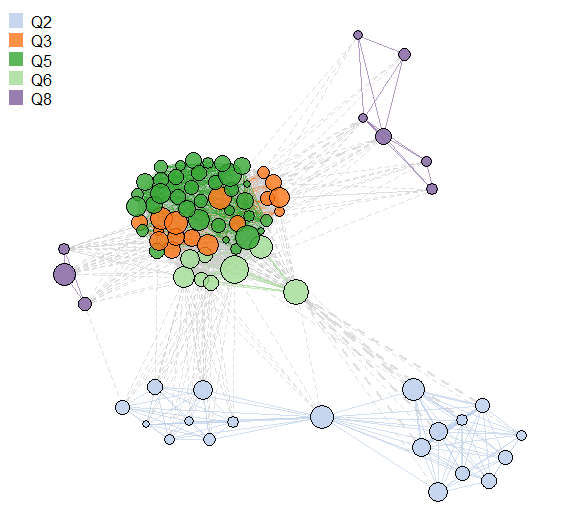}
    }
    \caption{Intra- and inter-community interaction intensities and community structure in the municipality network, 1985--1987.}
    \label{fig:cluster_interactions_II}
\end{figure}

Evidence of this expansion is the addition of 129 municipalities to the network, more than two thirds of the 184 municipalities observed in the previous period. Although violence did not persist in all previously affected municipalities, with 41 becoming inactive, many newly affected areas emerged in the same subregions, consistent with persistent proximity effects. As confrontation intensified, the municipal network developed more complex connections and the SBM detects nine communities, see Figure \ref{fig:cluster_interactions_II}. The resulting structure shows a denser center, thinner and more unequal links, and peripheral sets that either cluster locally or act as weak connectors rather than cohesive subregions.

In the third period, both the average within-community intensity \(\bar{\lambda}_i\) and the average between community intensity \(\bar{\lambda}_{ij}\) decline relative to 1982 to 1984, while the dispersion of \(\lambda_{ij}\) increases sharply. This implies that cohesion persists only for a subset of blocks and that links are sparser and far more heterogeneous, with a few comparatively strong channels standing out against many weak ties. The 227 municipalities are distributed across nine communities, \(Q_4\) with 74 municipalities and 32.6\%, \(Q_5\) with 34 and 15.0\%, \(Q_1\) with 24 and 10.6\%, \(Q_7\) with 23 and 10.1\%, \(Q_9\) with 20 and 8.8\%, \(Q_2\) with 19 and 8.4\%, \(Q_3\) with 16 and 7.0\%, \(Q_8\) with 9 and 4.0\%, and \(Q_6\) with 8 and 3.5\%.

Some features observed previously persist, including a giant component with a concentrated center and peripheral sets, but the system reorganizes in three ways. First, the core differentiates into a triad formed by \(Q_5\), \(Q_3\), and \(Q_6\), with a more uneven internal structure. \(Q_5\) spreads across multiple regions with moderate \(\bar{\lambda}_i\), \(Q_3\) is tighter and heavier and marks the period’s most intensely contested zone, and \(Q_6\) becomes smaller while acting as the principal connector that links the core to most surrounding blocks. Second, the periphery bifurcates into compact but lightly connected sets with mid range \(\bar{\lambda}_i\) and few external ties, for example \(Q_2\) in \textit{Urabá}, and diffuse low intensity bridge sets split into small subgraphs that touch many classes weakly, for example \(Q_8\). Alongside them, a corridor class \(Q_7\) connects municipalities at the expansion margins, and a localized zone of contestation consolidates in the northeast, \(Q_9\).

Third, the dispersion of inter community weights increases, so cross class exchange depends more on a small set of multi actor connectors. At the periphery, the network expands and splits. The largest class, \(Q_4\), groups low degree municipalities newly drawn into the conflict. Many are linked to recent FARC deployments and sit weakly in the giant component, or remain in small disconnected components. Paramilitary presence is thinner and uneven, so within class degree and strength remain low and dispersed. A similar composition, but at higher connectivity, appears in \(Q_1\), where most actors are locally bounded yet geographically proximate, yielding average degree and strength about three times those of \(Q_4\). Two focal zones stand out, southern \textit{Tolima} and northern \textit{Huila} under Fronts XXV and XVII, and \textit{Caquetá} and \textit{Guaviare} under the \textit{Bloque Sur} and the \textit{Bloque Oriental}.

The late 1980s combined progress toward a new peace agreement with renewed escalation of the civil war. While the Barco administration negotiated demobilization and political participation with insurgent groups willing to disarm, the FARC Secretariat sustained the expansion set out in the Seventh Conference plan. This decision reflected a continued commitment to armed struggle and the escalation of violence against UP militants and former combatants targeted by right wing armed groups. Between 1988 and 1990, paramilitary attacks intensified and expanded rapidly, producing more victims than violence attributed to the FARC. Massacres became the most distinctive tactic and increased more sharply than selective killings, forced disappearances, forced recruitment, sexual violence, and kidnappings. At the same time, FARC activity remained more geographically widespread and exhibited sharper territorial expansion.

\begin{figure}[!htb]
    \centering
    \subfigure[Community interaction intensities, $\lambda_{ij}$.]{
        \includegraphics[scale=0.4]{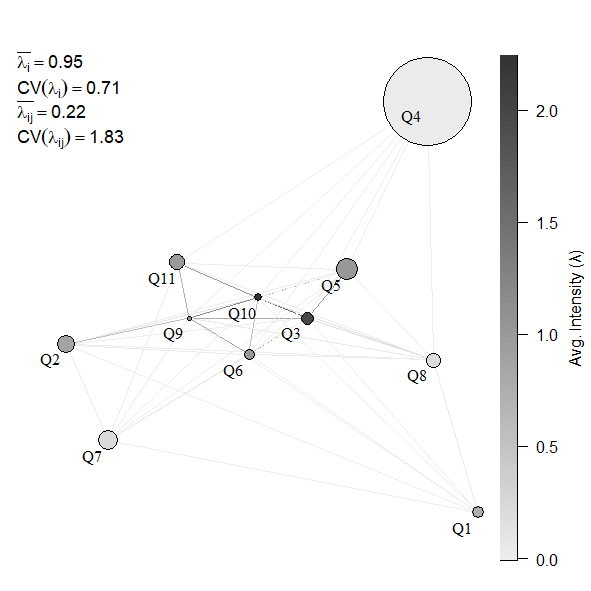}
    }
    \subfigure[Geographical distribution of communities.]{
        \includegraphics[scale=0.4]{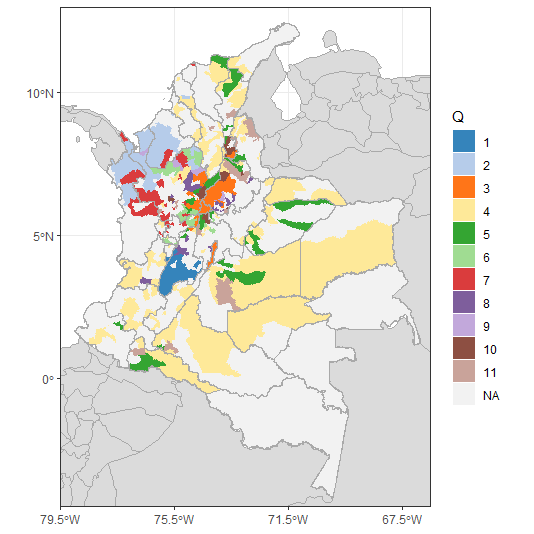}
    }
    \subfigure[Municipality network.]{
        \includegraphics[scale=0.4]{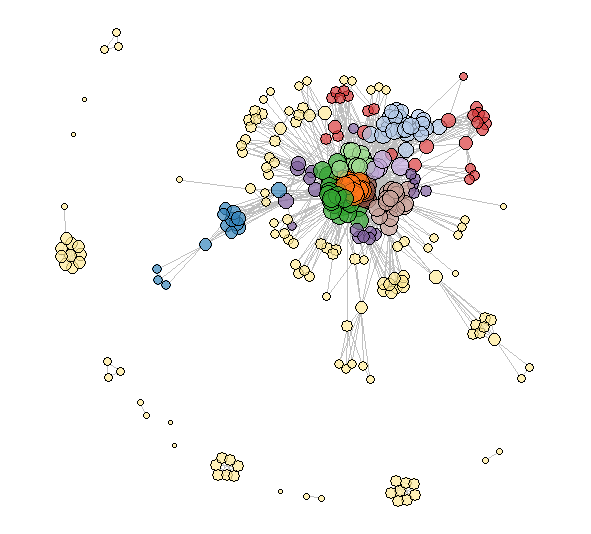}
    }
    \subfigure[Network core.]{
        \includegraphics[scale=0.4]{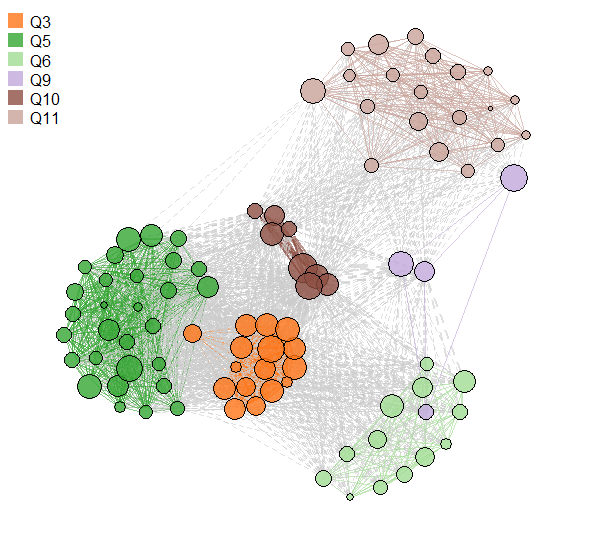}
    }
    \caption{Communities in the municipality network, 1988--1990.}
    \label{fig:cluster_interactions_IV}
\end{figure}

These changes unfolded unevenly across the territory. In 52.5\% of subregions, persistence, decline, and new incidence coexisted, indicating localized reconfiguration. In another 11.9\%, all previously affected municipalities continued to register events and adjacent municipalities reported first time attacks against civilians, which indicates expansion in coverage. A further 22.0\% correspond to areas with no prior incidents. In addition, 8.5\% show persistence without change, and 5.1\% combine persistence with subsequent decline. As in the previous period, these overlaps indicate that between 1988 and 1990 the conflict shifted and extended without necessarily receding from subregions that had already experienced violence. This yields three additional communities and changes in the composition of previous ones, see Figure \ref{fig:cluster_interactions_IV}.

The system displays a more fragmented and weight heterogeneous core, while preserving dense connectivity at the center and looser ties at the periphery. The partition includes eleven communities that distribute the 302 municipalities as follows. A single large class \(Q_4\) includes 128 municipalities, 42.4\%. An upper mid tier includes \(Q_7\) with 27 municipalities, 8.9\%, \(Q_5\) with 29, 9.6\%, and \(Q_1\) with 14, 4.6\%. The remaining communities are \(Q_2\) with 24, 7.9\%, \(Q_{11}\) with 20, 6.6\%, \(Q_3\) with 16, 5.3\%, \(Q_6\) with 13, 4.3\%, \(Q_8\) with 19, 6.3\%, \(Q_9\) with 4, 1.3\%, and \(Q_{10}\) with 8, 2.6\%. Consistent with this fragmentation, average within community edge weights range from about 1.00 in \(Q_7\) to 2.29 in \(Q_{10}\), indicating substantial variation in co-presence intensity across communities.

\newpage

At the core, overlap among expansive paramilitary formations, \textit{Masetos}, the structure led by \textit{Gonzalo} and \textit{Henry Pérez}, and \textit{Mano Negra}, organizes a set of tightly coupled blocks, though their interfaces are sharper and more unequal than in earlier periods. Community \(Q_{10}\) captures the triple overlap and links hotspots such as \textit{Puerto Berrío}, \textit{Barrancabermeja}, \textit{San Martín}, \textit{Aguachica}, and \textit{Yacopí}. Community \(Q_3\) captures \textit{Masetos} and Pérez intersections across \textit{Magdalena Medio}, spanning \textit{Antioquia}, \textit{Boyacá}, and \textit{Santander}, while \(Q_9\) maps Pérez and \textit{Mano Negra} contacts. By contrast, \(Q_5\) spreads \textit{Masetos} co-presence thinly across thirteen subregions, from the \textit{Sierra Nevada} to the \textit{Medio Putumayo}, rarely exceeding two municipalities per counterpart. Community \(Q_6\) concentrates ties between the Pérez group, FARC fronts, and smaller paramilitary labels along the \textit{Antioquia}, \textit{Cundinamarca}, and \textit{Eje Cafetero} hinge. Finally, \(Q_{11}\) delineates the \textit{Mano Negra} sector that does not overlap the core formations, extending across 12 subregions in 9 departments, with concentration in \textit{Santander}.

Two geographically clustered communities, \(Q_2\) and \(Q_1\), function as higher order subgraphs rather than single front cliques. In \(Q_2\), the \textit{Tangueros} interface with FARC fronts V and XVIII, extending the \textit{Atrato}, \textit{Bajo Cauca}, and northern \textit{Córdoba} chain into the \textit{Montes de María}. The cluster grows in size, maintains high within-community intensity, and increases outward connectivity as the projection in \textit{Urabá} intensifies. Clustering remains below one, which indicates overlapping pockets rather than universal co-presence. By contrast, \(Q_1\) in \textit{Tolima} remains a near clique under FARC XXI to XXV with the local \textit{Rojo Atá}. Transitivity is high, internal weights are uneven, and external ties concentrate through a small set of vertices, consistent with a consolidated guerrilla stronghold facing geographically constrained paramilitary reach.

A corridor configuration appears in \(Q_7\). It links FARC fronts LVII, XXXIV, XLVII, and IX, with XVIII at the margin, along the Cauca river axis across the \textit{Suroeste Antioqueño} and the \textit{Eje Cafetero}, with satellites in \textit{V Atrato}, \textit{Bajo Cauca}, and \textit{Norte}, \textit{Oriente}, and \textit{Occidente Antioqueño}. Under the Northwestern Bloc, neighboring fronts held contiguous areas, so cooccurrence remains boundary bound. Accordingly, within community edges lie near unit weight, \(w_{uv} \approx 1\), and the betweenness distribution is right skewed, since outward links disperse across multiple connectors linking \textit{Urabá} to interior subregions.

At the periphery, \(Q_4\) again aggregates small scattered components and non adjacent subgraphs across twenty five departments, territories where locally bounded actors intersect only marginally with broader networks, while a smaller set of nodes remains disconnected from the giant component. Degrees remain low both within and between communities. Overall, the 1988 to 1990 system preserves the giant component template, but it shifts the load onto a small number of multi actor interfaces, sharpening core fragmentation and making cross community traffic more selective and unequal.

In 1990, peace accords led to the demobilization of M-19, EPL, \textit{Quintín Lame}, and PRT, and a new constitution was enacted, while the FARC broke off talks. The period opened with the bombing of \textit{Casa Verde} and closed with the FARC Eighth Conference. In parallel, the \textit{Medellín} Cartel and paramilitary alliance in the \textit{Magdalena Medio} collapsed. \textit{Henry Pérez} was assassinated, some units partially demobilized, and failed reintegration produced autonomous successor structures. In \textit{Urabá}, as the EPL demobilized, \textit{Fidel Castaño} announced the dismantling of the \textit{Tangueros}, but instead expanded by recruiting former EPL combatants and waging war against \textit{Escobar} through December 1993.

The SBM detects fifteen connectivity based clusters, see Figure \ref{fig:sankey_municipalities}, and indicates a shift toward a more community oriented organization, with stronger interiors and more selective interfaces. Within community interaction intensities increase in both level and spread, while between community ties average about one eighth of the mean within-community intensity. The distribution is highly skewed, most \(\lambda_{ij}\) values are small and a limited subset is large. Municipalities are also more evenly distributed across classes.

The former \textit{Magdalena Medio} hub splits into distinct roles. One part becomes a corridor controlled by the \textit{Autodefensas del Magdalena Medio} led by \textit{Ramón Isaza}, \(Q_8\), covering the southern and central \textit{Magdalena Medio} and the connection between \textit{La Dorada} and \textit{Honda}. Another part becomes a contested overlap zone, \(Q_3\), where \textit{Isaza}’s group intermittently advances into the \textit{Puerto Boyacá} basin amid the violent reordering after the assassination of \textit{Henry Pérez}, and tension increases rather than declines. A third part forms a successor rim in northwestern \textit{Cundinamarca}, the \textit{Yacopí}, \textit{La Palma}, \textit{Topaipí}, \textit{Caparrapí}, and \textit{Puerto Salgar} sector, where the \textit{Autodefensas Campesinas de Yacopí} take control of the southwest \textit{Pérez} and \textit{Gacha} area. This explains why the former core no longer functions as a single entity, the mass remains, but it is divided into a stable corridor, a high tension overlap, and a new rim with separate external connections.

\begin{figure}[!htb]
    \centering
    \subfigure[Community interaction intensities, $\lambda_{ij}$.]{
        \includegraphics[scale=0.4]{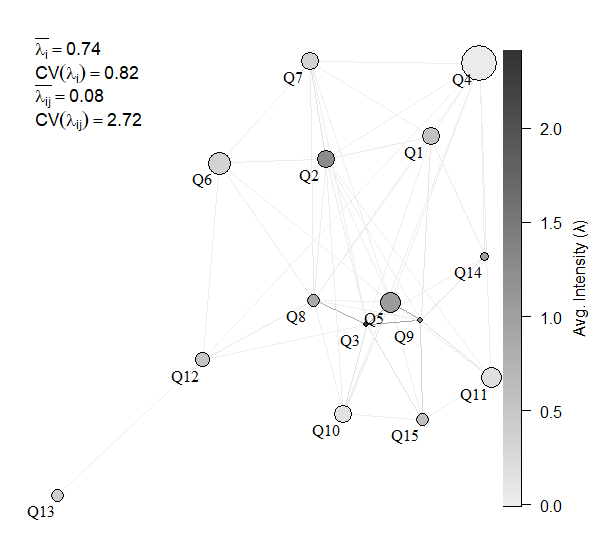}
    }
    \subfigure[Geographical distribution of communities.]{
        \includegraphics[scale=0.4]{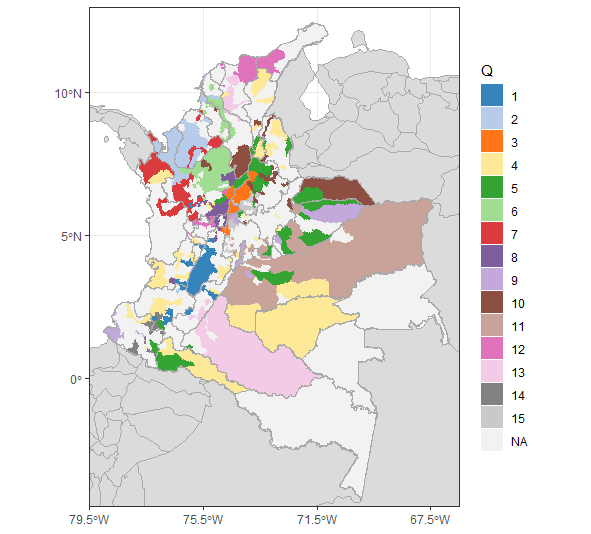}
    }
    \subfigure[Municipality network.]{
        \includegraphics[scale=0.4]{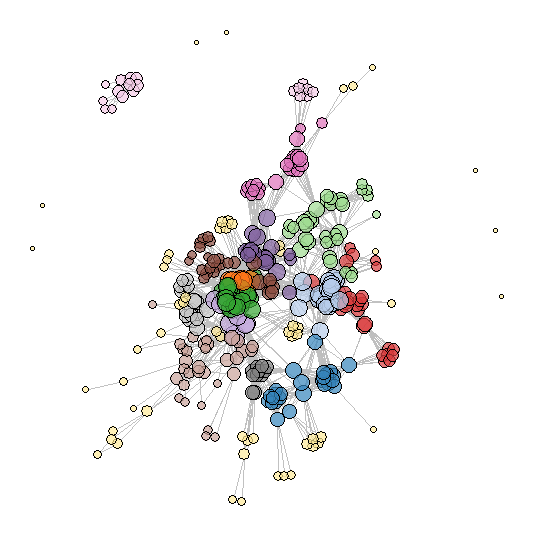}
    }
    \subfigure[$Q_3$ ego graph, first-order structure.]{
        \includegraphics[scale=0.4]{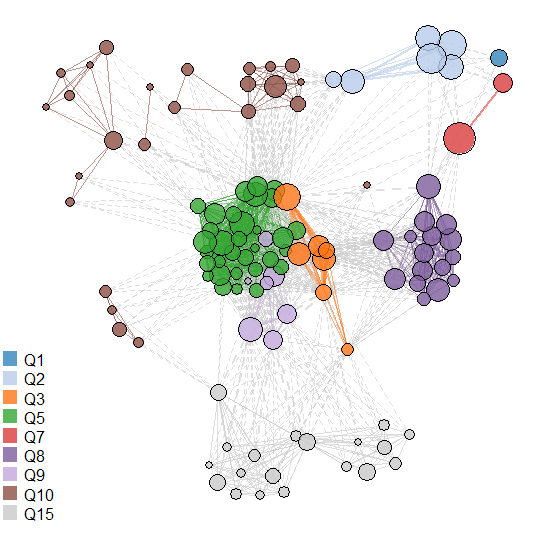}
    }
    \caption{Communities in the municipality network, 1991--1993.}
    \label{fig:cluster_interactions_V}
\end{figure}

Eastward, the system organizes as a bifurcating chain. Community \(Q_{11}\), closely coupled to \(Q_9\), links the middle \textit{Magdalena} to the \textit{Piedemonte Llanero} and then splits. One branch follows the \textit{Ariari} and \textit{Guayabero} corridor toward southern \textit{Meta}. The other runs along the upper \textit{Meta} plain toward the \textit{Altillanura}, with a lateral spur into \textit{Casanare}. Externally, \(Q_{11}\) connects to five partners but concentrates most outward weight on two, primarily \(Q_9\). This matches a structure in which fourteen FARC fronts under the \textit{Bloque Oriental} meet three paramilitary poles, \textit{Autodefensas de San Martín}, Carranza’s security apparatus, and the \textit{Autodefensas Campesinas del Casanare}. The resulting geometry is a selective trunk with two heavy attachments and weaker residual contacts elsewhere.

Most \textit{Masetos} tagged municipalities fall in \(Q_5\). Its relatively high internal degree, with mean \(\approx 48.1\), and low dispersion indicate a domain whose ties concentrate within rather than outward. Community \(Q_9\) sits between \(Q_5\) and the wider system. It connects to seven partner classes but allocates about two thirds of its external weight to two main counterparts, combining two heavy interfaces with a long tail of weak links. Alongside, \(Q_{10}\) displays a co-presence pattern similar to \(Q_5\), except for the absence of the \textit{Masetos}. Its within community degree distribution is uneven, with clustering \(\approx 0.91\), mean within degree \(\approx 4.4\), and \(\mathrm{CV} \approx 55\%\), suggesting a union of several small locally complete subgraphs with limited overlap. Two partners, \(Q_5\) and \(Q_3\), carry about 81\% of its outward weight. The remaining links to \(Q_2\), \(Q_4\), \(Q_8\), and \(Q_{15}\) are residual.

On the FARC dominated side, \(Q_{14}\) includes municipalities in \textit{Nariño} within the Front XXIX area of operations and forms a segmented chain along Andean passes. Within community ties are near complete, while outward links are numerous but uniformly light, with effective partners \(\approx 7.6\), which indicates diffuse contact rather than a few heavy overlaps. Community \(Q_1\), where Fronts VI and XXI were active, forms a compact two segment corridor into \textit{Tolima}. It touches many partners but concentrates most outward weight on a small subset, with effective partners \(\approx 2.2\). Community \(Q_{15}\), centered in northwest \textit{Cundinamarca}, is similar in size to \(Q_{13}\) but functions as a leaf, connected to the rest of the network almost entirely through its tie to \(Q_9\). Community \(Q_{13}\) comprises two non adjacent subgraphs, one linking \textit{Caquetá} and \textit{Huila} through vertices outside the giant component. These municipalities lie in an area disputed by Fronts III, XIII, and XIV, and by the mobile column \textit{Teófilo Forero}. After the aerial bombardment of \textit{Casa Verde}, these patterns indicate renewed concentration on boundary limited multi actor overlap, contiguous control, and durable strongholds.

The \textit{Urabá} margin appears as a paired configuration with diverging roles. In \(Q_2\), the \textit{Tangueros} consolidate across the twenty six municipalities of \textit{Córdoba} and the \textit{Urabá antioqueño} through post-demobilization recruitment from the EPL. Externally, ties narrow and concentrate, with effective partners \(\approx 2.49\) and top two share \(\approx 0.94\), reflecting fewer and heavier channels as links to \textit{Magdalena Medio} weaken. Community \(Q_7\) remains boundary bound. Internal degree and clustering decline, and outward links disperse across multiple connectors, consistent with neighboring FARC fronts holding contiguous areas and meeting the \textit{Tangueros} at shared borders. Several of those fronts, V, XVIII, XXXIV, and LVII, appear in both \(Q_2\) and \(Q_7\), underscoring a contact zone shaped by adjacency rather than deep overlap.

A final corridor encloses the northern belt. Community \(Q_6\) runs from the \textit{Bajo Cauca} and \textit{Norte de Antioquia} junctions into the \textit{Montes de María}. On the Antioquia side, road river hinges such as \textit{Zaragoza}, \textit{El Bagre}, \textit{Caucasia}, \textit{Tarazá}, and \textit{Yarumal}, \textit{Valdivia}, \textit{Ituango}, \textit{Anorí} host locally anchored self defence labels, including ANA and \textit{Doce Apóstoles}, tied to the 1980s nexus of private security, political brokers, and narcotrafficking finance. Eastward, the corridor crosses the \textit{La Mojana} and \textit{Canal del Dique} hinge to an interior Caribbean segment where FARC fronts XXXV and XXXVII, and on the Antioquia flank XXXVI and XVIII, meet local counterinsurgency labels. The result is a connector with shared boundaries to the \textit{Tangueros} zone \(Q_2\) and the \textit{Magdalena Medio} corridor \(Q_8\), and lighter links to \(Q_5\) and \(Q_7\).

At the periphery, \(Q_4\) preserves the configuration of multiple small non contiguous components. High clustering coexists with low mean within community degree and strong dispersion, with clustering \(\approx 0.92\), mean within degree \(\approx 2.18\), and \(\mathrm{CV} \approx 71\%\). It connects to nine classes with low concentration, with top two share \(\approx 0.52\). Heavier contact points to \(Q_5\), with additional lighter links to \(Q_1\), \(Q_{10}\), \(Q_8\), \(Q_{11}\), \(Q_7\), and \(Q_{14}\). Composition aligns with this structure, numerous FARC fronts from several blocs and a smaller set of non FARC actors, all lightly integrated into systemwide exchange.

After 1993, paramilitary expansion shifted from the \textit{Tangueros} to the \textit{Autodefensas Cam\-pe\-si\-nas de Córdoba y Urabá}, ACCU, while the FARC implemented the Eighth Conference adjustments, including bloc based command, offensive pushes toward urban corridors, and front doubling, most clearly in the \textit{Llanos Orientales}. The partition contracts to twelve classes, but connectivity thickens along a small set of corridors that carry a disproportionate share of between community exchange.

Community \(Q_2\) concentrates the ACCU consolidation corridor from \textit{Urabá} into the Antioquia Caribbean hinge and outward along the coastal belt. Municipalities that were dispersed across four classes in 1991 to 1993 cohere into a single dominant block whose internal graph is near complete with unit weight ties. Co-presence is pervasive and largely single locality, layered along successive advances from the banana axis into \textit{Norte} and \textit{Occidente Antioqueño}, then \textit{Suroeste}, \textit{Nordeste}, and \textit{Oriente}, and through \textit{Bajo Cauca} toward the \textit{La Mojana} connector. Exposure to neighbors is wide, with degree far above the within community benchmark, but concurrence with counterpart structures is typically thin, consistent with contact at boundaries rather than sustained multi municipal overlap.

\begin{figure}[!htb]
    \centering
    \subfigure[Community interaction intensities, $\lambda_{ij}$.]{
        \includegraphics[scale=0.33]{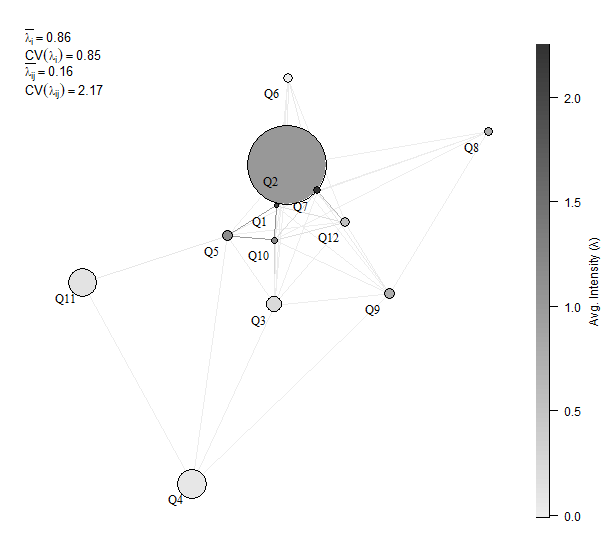}
    }
    \subfigure[Geographical distribution of communities.]{
        \includegraphics[scale=0.48]{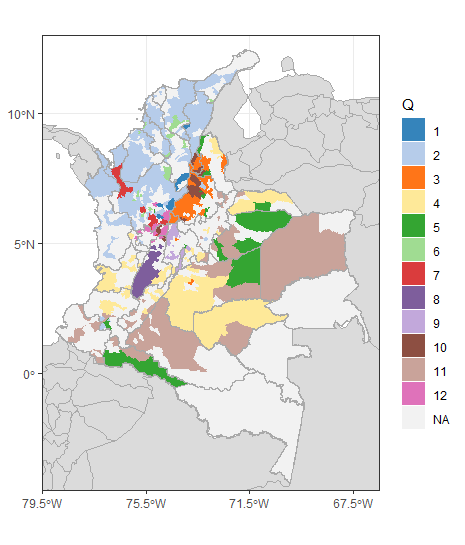}
    }
    \subfigure[Municipality network.]{
        \includegraphics[scale=0.33]{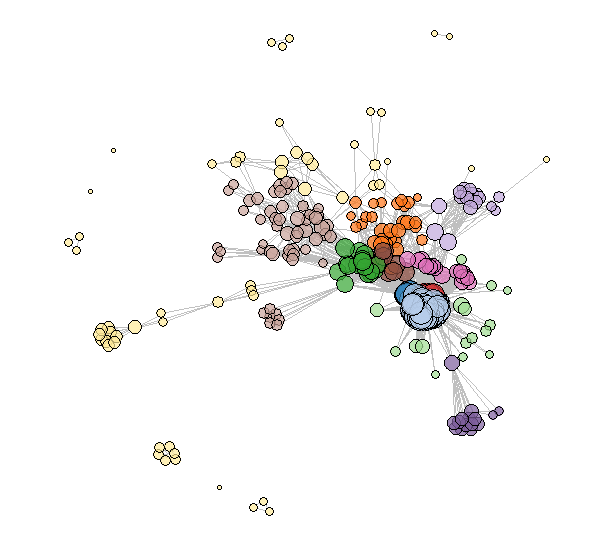}
    }
    \subfigure[$Q_2$ ego graph, first-order structure.]{
        \includegraphics[scale=0.33]{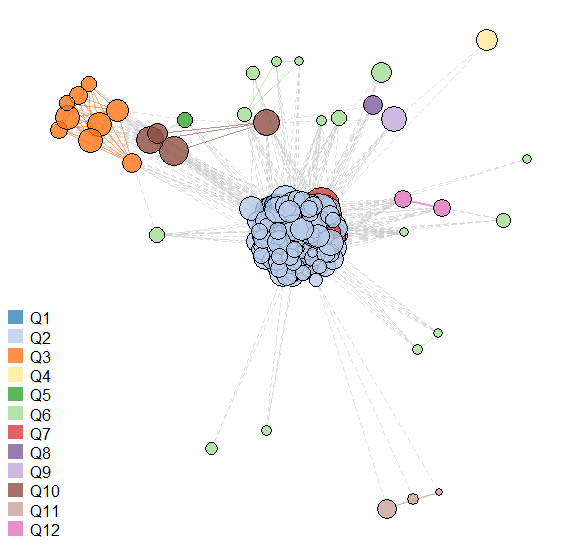}
    }
    \caption{Communities in the municipality network, 1994--1996.}
    \label{fig:cluster_interactions_VI}
\end{figure}

A tighter ACCU overlap zone appears in \(Q_7\), with recurrent joint ACCU and FARC presence. Across a limited set of municipalities in \textit{Urabá}, \textit{Valle de Aburrá}, and the \textit{Oriente Antioqueño}, at least two units operate in every locality and each structure spans several municipalities on average. This indicates sustained ACCU and FARC engagement under the \textit{Bloque Noroccidental}. Community \(Q_7\) links the broad coverage of \(Q_2\) to the western bank corridor in \(Q_{12}\), forming a west center hinge that channels traffic into the \textit{Magdalena Medio} basin.

The 1994 to 1996 partition deepens the post 1991 split in the \textit{Magdalena Medio}. Community \(Q_{12}\) maps the \textit{Isaza} side of the basin, with few external ties and a northwest orientation, where Fronts XLVII and IX contest local control. Across the river, \(Q_3\) traces a larger and more exposed northward corridor as the \textit{Puerto Boyacá} remnant advances through \textit{Santander} toward southern \textit{Cesar}. It intersects circuits associated with \textit{Camilo Morantes} and \textit{Juancho} Prada and maintains a strong attachment to the central hinge in \(Q_{10}\). By contrast, \(Q_9\) is a compact and highly cohesive FARC set in northwestern \textit{Cundinamarca}, where multiple \textit{Bloque Oriental} fronts face the \textit{Autodefensas Campesinas de Yacopí}. Its signature is dense internal transitivity with selective external contact.

As in 1991 to 1993, two adjacent communities, \(Q_{10}\) and \(Q_5\), concentrate most between-community weight around the triad \(Q_3\), \(Q_9\), and \(Q_{12}\). The \(Q_{10}\) and \(Q_5\) axis anchors the MAS zone and concentrates local between community traffic that links fragmented \textit{Magdalena Medio} tracts to neighboring regions. Community \(Q_{10}\) channels much of that interchange into \(Q_5\). Community \(Q_5\) then projects eastward via \(Q_{11}\) along the \textit{Cordillera Oriental} line toward \textit{Bogotá}, consistent with Eighth Conference directives. The result is a persistent chain of linkages connecting paramilitary dominated piedmont tracts and the \textit{Llanos Orientales} to municipalities with active FARC fronts in \textit{Meta}, \textit{Vichada}, \textit{Casanare}, and \textit{Caquetá}.

Peripheral vertices continue to aggregate in a single class, \(Q_4\), composed of small components and minor subgraphs attached to the giant component through a limited set of bridges. In the 1994 to 1996 network, with total edges \(= 17{,}043\), community \(Q_4\) includes 154 connections, of which 28 are cut edges. This architecture places most of the system load on the central and northwest axis while preserving a wide but lightly integrated periphery.

We now examine connection patterns in 1997 to 1999. This period marks the transition from dispersed paramilitary coalitions to a federated structure, the \textit{Autodefensas Unidas de Colombia}, coordinated under a single command while remaining decentralized across subregional blocs. With this reorganization, paramilitary forces moved from localized defense to coordinated operations along new corridors and launched anti guerrilla incursions into previously unentered areas. Paramilitary groups no longer acted in isolation. They conducted offensives beyond their traditional zones, linking regions and expanding eastward and southward. Meanwhile, the FARC pursued a three part strategy. It concentrated roughly half of its force along the \textit{Cordillera Oriental} and continued the advance toward \textit{Bogotá}. It created institutional vacuums by seizing municipal seats, forcing police withdrawals, and disrupting elections. It also formed mobile units for short assaults designed to overrun positions and withdraw before counteraction. This expansion was sustained by increases in kidnapping and recruitment and by struggles over drug trafficking routes and key nodes.

\begin{figure}[!htb]
    \centering
    \subfigure[Community interaction intensities, $\lambda_{ij}$.]{
        \includegraphics[scale=0.4]{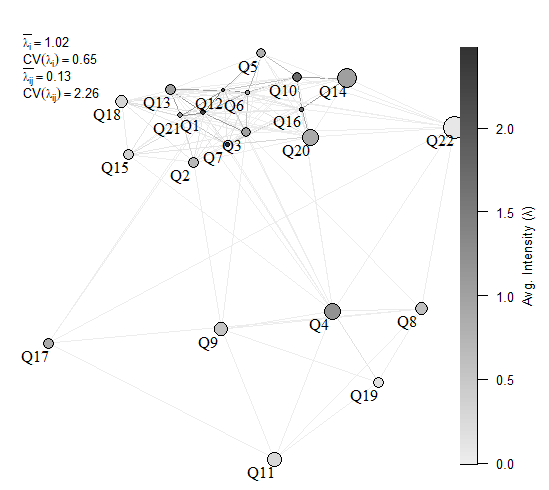}
    }
    \subfigure[Geographical distribution of communities.]{
        \includegraphics[scale=0.4]{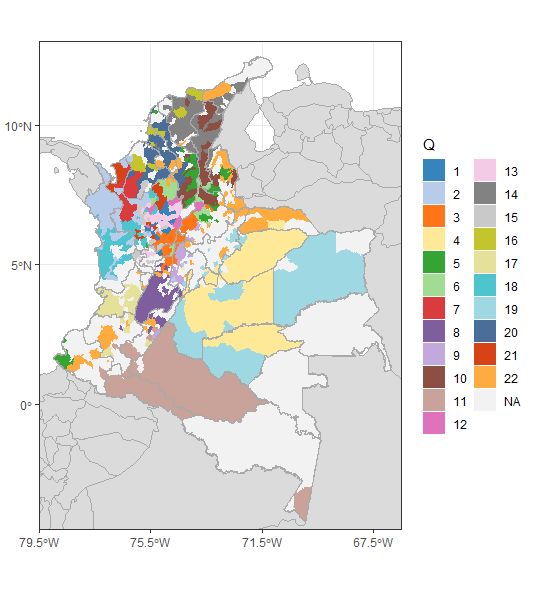}
    }
    \subfigure[Municipality network.]{
        \includegraphics[scale=0.4]{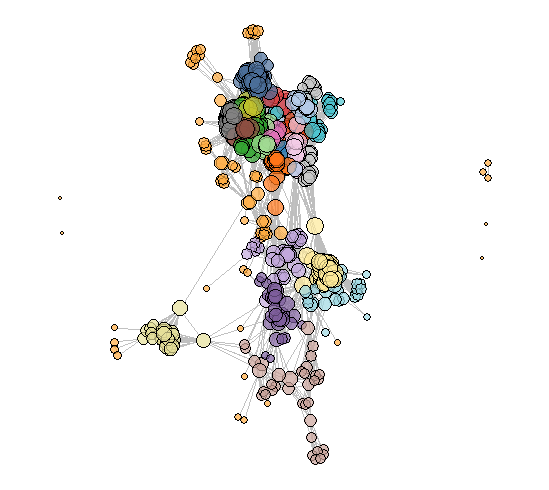}
    }
    \subfigure[$Q_3$ ego graph, first-order structure.]{
        \includegraphics[scale=0.4]{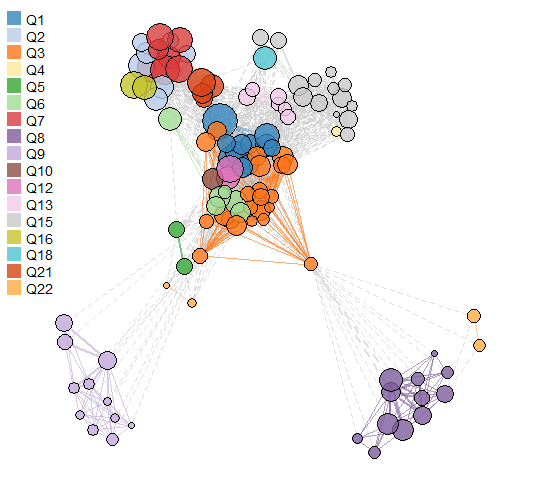}
    }
    \caption{Communities in the municipality network, 1997--1999.}
    \label{fig:cluster_interactions_VII}
\end{figure}

In this setting, interactions among armed structures yield a partition into 22 communities in 1997 to 1999, see Figure \ref{fig:cluster_interactions_VII}. The size distribution includes three larger sets with 65, 49, and 41 vertices, which together account for 27.7\%. It then includes one upper mid set with 37 vertices, 6.6\%, a middle band of twelve communities with 21 to 33 vertices, 55.7\%, and a six community tail with at most 17 vertices, 10.0\%. The system exhibits a dual core configuration, one dense and one looser, connected by a small number of weak ties. Average within-community intensity remains high, with \(\bar{\lambda}_i = 1\). By contrast, between community connectivity decreases and becomes highly heterogeneous, with \(\bar{\lambda}_{ij} = 0.14\) and \(\mathrm{CV}[\lambda_{ij}] = 2.18\). This indicates a heavy tailed pattern in which a small set of edges carries most of the weight. Cross regional connectivity therefore relies on a limited number of low weight bridges concentrated on high betweenness interface communities, consistent with selective coupling rather than system wide cohesion.

We study these structures through two lenses. The first is coverage concentration. For each actor, we track its local coverage within a class, \(r\), and its territorial focus across classes, \(f\), and combine them as \(D = \sqrt{r f}\). Classes dominated by the same actor and connected by strong between community ties are treated as cohesive higher order regions. These sets exhibit strong links driven by the joint presence of highly influential actors that both reach a substantial share of municipalities within each block and concentrate a large fraction of their overall activity there, with \(D \ge 0.25\). The second lens is within community co-presence. When multiple dominant actors share municipalities, the projection generates intertwined subgraphs that reflect mixed or competitive control. We use these patterns to distinguish actor anchored regions from mixed corridors.

Relative to 1994 to 1996, two shifts define 1997 to 1999. First, the west to east paramilitary spine densifies as the AUC consolidates twenty four armed substructures, twenty of them operating across the central and northern belt where they overlap nineteen FARC fronts under the \textit{Bloque Noroccidental}, \textit{Bloque Caribe}, and \textit{Bloque Magdalena Medio}. These interactions induce the densest zone of the giant component, and the SBM identifies eight tightly connected communities that in the previous period belonged to the ACCU anchored cluster, see Figure \ref{fig:cluster_interactions_VII}. Second, between community exchange becomes more unequal. While \(\bar{\lambda}_{ij}\) declines, dispersion rises, placing most cross community flow on a small number of interfaces.

Within this densified belt, the strongest predicted ties form a triad \(Q_{10}\), \(Q_{14}\), and \(Q_{16}\) associated with the \textit{Bloque Norte}. Community \(Q_{14}\) spans a broad operating area with modest mean edge weights, \(\overline{w} \approx 1.04\), consistent with a single dominant presence with few additional overlaps. Community \(Q_{16}\) groups a small spatially non adjacent set with near two expected cooccurrence, capturing joint activity by substructures backed by \textit{Salvatore Mancuso} at the hinge between \textit{Bloque Córdoba} and \textit{Montes de María} and the \textit{Bloque Norte}. Community \(Q_{10}\) functions as a medium scale hub linking this northern core to the \textit{Bloque Central Bolívar} and to \textit{Autodefensas Unidas de Santander y Sur del Cesar}. The predicted between community intensity within this triangle, about \(1.7\), indicates dense multi actor presence. The point is structural, a small number of interfaces concentrate most cross community exchange along the \textit{San Lucas}, \textit{Magdalena}, and \textit{Cauca} axis and toward its coastal extensions.

A second nucleus centers on the \textit{Bloque Central Bolívar} across \(Q_6\), \(Q_5\), and \(Q_{12}\), each distinguished by its secondary overlaps. For these communities, the top two external links carry 43 to 70\% of between-community weight, with the remainder spread thinly across additional partners, with five to nine effective counterparts. Community \(Q_6\) is the most evenly outward connected, sustaining dense but non dominant exchange with at least seven neighbors around \textit{Bajo Cauca}, the BCB operational epicenter. Community \(Q_5\) maps an expansion corridor with similar but milder reach. Community \(Q_{12}\) is the heaviest corridor. It links the BCB to the \textit{Bloque Metro} footprint and exhibits the highest between community strength among BCB adjacent sets. It transfers flow between the northeast and east hinge and the \textit{Eje Cafetero} without relying on a single bridge.

The \textit{Bloque Metro} footprint is split across \(Q_1\), \(Q_{13}\), and \(Q_{21}\). These represent, respectively, the northeast and east hinge with key interfaces beyond \textit{Metro} embedded areas, the belt around the \textit{Valle de Aburrá} with fewer overlaps, and the southwestern corridor into the \textit{Eje Cafetero} that acts as the principal southbound bridge. In \textit{Urabá}, the subsystem formed by \(Q_2\) and \(Q_7\) records the contest between the \textit{Bloque Élmer Cárdenas} and \textit{Bloque Bananero} and the FARC \textit{Bloque Noroccidental}. Both communities overlap with Front~LVII, and \(Q_7\) also intersects Fronts~V, XXXV, and LVIII. They differ in their connectivity patterns. Community \(Q_2\) shows moderate and relatively even overlap, with mean edge weight \(\bar{w} = 1.07\) and \(\mathrm{CV} = 24.6\), and operates as a feeder toward the broader northwestern front, especially \(Q_{15}\). Community \(Q_7\) is a multi actor junction with stronger and more uneven ties, with mean \(= 2.63\) and \(\mathrm{CV} = 53.9\), toward \textit{La Mojana}, the BCB, \textit{Metro}, and \textit{Magdalena Medio}. This asymmetry concentrates exchange on \(Q_7\) and positions the banana corridor and the \textit{Nudo del Paramillo} as a principal interface.

Eastward, paramilitary expansion from \textit{Urabá} links discontinuous zones in western \textit{Cundinamarca} and the \textit{Llanos Orientales}. The \textit{Bloque Élmer Cárdenas} connects \(Q_2\) to \(Q_9\) along the central Andean corridor, while the \textit{Bloque Centauros} dominates \(Q_4\) across more than forty municipalities in \textit{Meta}, with projections toward \textit{Casanare} and \textit{Guaviare}. Along the same corridor, \(Q_9\) remains guerrilla led, with Fronts~XXII and XLII, while \(Q_4\) reverses the balance, with AUC present throughout and FARC present in about half of the set. On its rim, \(Q_{19}\) isolates a low intensity perimeter, with moderate clustering at \(0.83\), thin internal connectivity, and limited external exchange, consistent with a boundary strip rather than a consolidated extension.

Communities \(Q_4\) through \(Q_{19}\) mark the edge of the dense northern and central subsystem and a broader belt of weaker guerrilla led connectivity. Adjacent to the core, predicted between community intensities delineate a second macro region of loosely connected sets, \(Q_8\), \(Q_9\), \(Q_{11}\), \(Q_{17}\), \(Q_{19}\), and \(Q_{22}\). Overall degrees remain small, about 7 to 20, versus values above 70 in the core paramilitary corridors. Within community heterogeneity is high, with \(\mathrm{CV} \approx 35\%\) to \(95\%\), and clustering is moderate at about \(0.8\), with \(Q_{11} = 0.64\). These characteristics indicate territorial segmentation rather than multi actor overlap across the southern and eastern periphery, placing these classes in the lower tail of between community strength.

In 2000 to 2001, community sizes are moderately right skewed across 27 blocks. Sizes range from 2 to 67 municipalities. The median is 24, and the mean is about 25.4. The five largest blocks, with at least 42 vertices each, account for about 38.5\% of municipalities, while the eight smallest blocks, with at most 11, account for about 10.5\%. Most communities fall in a mid size band, while a small upper tail accounts for a substantial share of territorial coverage and therefore exerts greater leverage on any reweighting of between community ties.

During the eighth period, the two module configuration linked by a limited set of high betweenness hinges reorganizes into a core in which intermediation is shared across several blocks. The \(k\) core number increases from 77 to 111, meaning more nodes satisfy the minimum degree threshold and dense neighborhoods cover a larger share of the graph. In parallel, the largest biconnected component covers 668 of 685 municipalities, or 97.5\%, and the number of articulation points declines. Consistently, the normalized node betweenness distribution shifts downward in its center, retains a similar ceiling in its upper tail, and shows reduced dispersion. High \(\bar{\lambda}_i\) core classes connect more often to mid \(\bar{\lambda}_i\) neighbors, while several southern and eastern satellites gain partners without becoming hubs.

\begin{figure}[!htb]
    \centering
    \subfigure[Community interaction intensities, $\lambda_{ij}$.]{
        \includegraphics[scale=0.33]{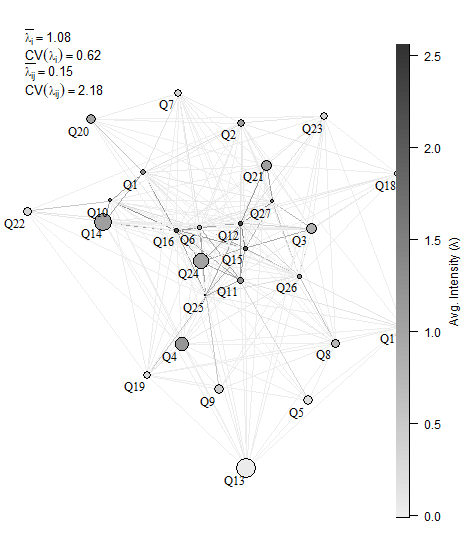}
    }
    \subfigure[Geographical distribution of communities.]{
        \includegraphics[scale=0.48]{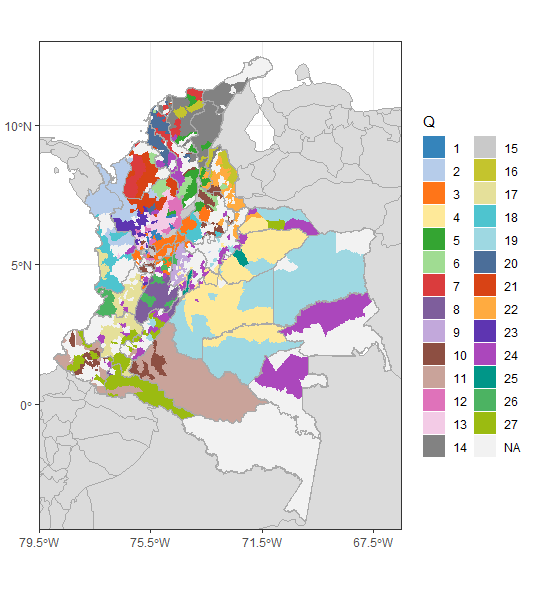}
    }
    \subfigure[Municipality network.]{
        \includegraphics[scale=0.33]{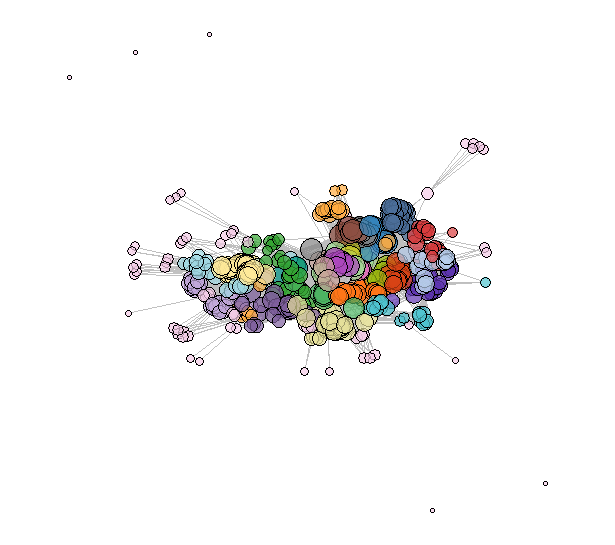}
    }
    \subfigure[$Q_{25}$ ego graph, first-order structure.]{
        \includegraphics[scale=0.48]{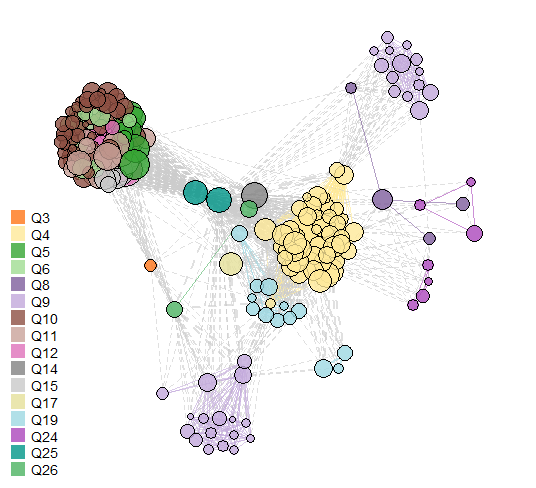}
    }
    \caption{Communities in the municipality network, 2000--2001.}
    \label{fig:cluster_interactions_VIII}
\end{figure}

As alternative paths multiply, the sharp divide separating the two cores in the previous period begins to soften. The SBM redraws central boundaries. Municipalities near former hinges are absorbed into a consolidated center, satellites attached to this center are reassigned around it, and blocks that previously acted as connectors begin to resemble ordinary core components. The result is a tighter central fabric in which intermediation is shared across multiple hubs rather than concentrated on a small set of interfaces. This reorganization helps explain why cross period one to one correspondences with earlier corridors weaken.

Within each community \(Q\), we assess whether cohesion is anchor led or mixed by combining a dominance indicator with two composition metrics. Let \(V_Q\) denote the set of municipalities in \(Q\), with \(M = |V_Q|\), and let \(A_Q\) denote the set of armed structures active in \(Q\). For each actor \(i \in A_Q\), let \(m_i\) be the number of municipalities in which \(i\) is present during the period. We summarize concentration using two measures. The first is top actor coverage, \(\mathrm{cov}_{\max}(Q) = \max_i m_i / M\), the share of municipalities covered by the most widespread actor in \(Q\). The second is the Herfindahl Hirschman index computed on presence shares, \(\mathrm{HHI}(Q) = \sum_{i \in A_Q} p_i^2\), where \(p_i = m_i / T\) and \(T = \sum_{j \in A_Q} m_j\) is the total number of municipality actor incidences in \(Q\). To evaluate mixture, we estimate Pielou’s evenness \(J(Q) = H(Q) / \log S\), where \(H(Q) = -\sum_{i \in A_Q} p_i \log p_i\) and \(S = |\{i \in A_Q \mid m_i > 0\}|\). High \(\mathrm{cov}_{\max}\) and high \(\mathrm{HHI}\) indicate dominance by one or a small set of actors. Values of \(J\) near 1 indicate balanced co-presence, while \(J\) near 0 indicates highly uneven mixtures.

For classes where the top actor covers less than 0.4 of municipalities, \(Q_{23}\), \(Q_{24}\), and \(Q_{27}\), incidence concentration is low, with median \(\mathrm{HHI} \approx 0.096\), and evenness is high, with median \(J_{\mathrm{even}} \approx 0.907\). This indicates genuinely mixed compositions in which no single organization shapes most within class ties. How this mixture is routed varies. In \(Q_{23}\), the mix is clustered in western \textit{Antioquia}, but cross group contact is funneled through a small set of high betweenness municipalities, producing a segmented corridor like pattern. Community \(Q_{27}\) also shows a right skewed betweenness distribution, but the area leans toward rebel influence. Fourteen FARC fronts appear with limited paramilitary contact across \textit{Nariño}, \textit{Cauca}, and \textit{Huila}, so routes mainly run within FARC territories through a few gates. By contrast, \(Q_{24}\) is diffuse and non contiguous. Scattered micro components tie lightly to many neighbors across regions, with non trivial links to eleven communities, surrounding localized contestation belts, for example around \(Q_4\) and \(Q_{19}\), and functioning as a broad peripheral wrapper and weak connector.

As top actor coverage rises into the \(0.4 \le \mathrm{cov} < 0.5\) band, incidence concentration more than doubles at the median, with \(\mathrm{HHI} \approx 0.249\), while evenness declines, with \(J_{\mathrm{even}} \approx 0.852\). This is consistent with the emergence of anchor led mixtures rather than balanced co-presence. Communities in this band, \(Q_{18}\), \(Q_{19}\), \(Q_{21}\), and \(Q_{22}\), are similar in size, about 18 to 25 municipalities, show moderate actor richness, with roughly six armed structures active, and exhibit overlaps that typically involve one to two actors per locality. Within class ties are organized by a primary actor appearing in close to half of the municipalities, with secondary actors recurring in patterned pairings.

In the broad mid dominance band, \(0.5 \le \mathrm{cov} < 0.8\), concentration stabilizes at a moderately high level, with median \(\mathrm{HHI} \approx 0.24\), while evenness declines, with median \(J_{\mathrm{even}} \approx 0.72\). This indicates classes structured by two, and sometimes three, leading organizations rather than a single anchor. Conflict expression in these blocks is best read as dual leadership mixing. Within class ties are dense but braided, with repeated co-presence of the same two actors across many municipalities and a small set of recurrent secondary actors. Outward exchange is selective but not chokepointed. Between class weight is distributed across multiple partners, supporting parallel channels rather than a single bridge. In \(Q_8\), centered in \textit{Tolima}, internal connectivity is sustained by overlaps of the two lead actors, \textit{Bloque Tolima} and FARC Front XXI, with Front XXV reinforcing co-presence, raising within class edge weights and adding redundant paths. Clustering slightly above 0.8 indicates incomplete coverage across vertices, but overlap pockets are well linked, while outward ties remain dispersed among mid strength partners. By contrast, \(Q_9\) resolves into two compact non contiguous cliques, one in northwestern \textit{Cundinamarca}, tied to FARC Front XXII, and another in the southwest, tied to Front XLII, connected by a light pendant. With only six partners and top two share \(\approx 0.62\), between class exchange concentrates on a limited set of edges. A third case, \(Q_3\), preserves dense dual actor mixing internally but concentrates outward flow more sharply, with top two share \(\approx 0.52\), yielding a block that is internally interwoven and externally oriented toward a narrow set of counterparts. Overall, we observe multi actor dominance within, coupled with controlled connectivity outward across multiple paths.

At very high dominance, \(\mathrm{cov} \ge 0.8\), the central tendency of incidence concentration increases, but dispersion widens substantially, with median \(\approx 0.168\), mean \(\approx 0.252\), and maximum \(\approx 0.774\). Evenness spans a wide range, \(J_{\mathrm{even}} \in [0.309, 0.975]\). This heterogeneity indicates two configurations. In the first, the dominant actor rarely encounters other armed structures, yielding near monopoly coverage. In the second, two or more leading actors maintain substantial coverage, and additional secondary actors overlap with them. The first group includes \(Q_{10}\), \(Q_{14}\), and \(Q_{20}\). The latter two correspond to consolidating areas of the \textit{Bloque Norte} and the \textit{Bloque Héroes de los Montes de María}, while \(Q_{10}\) comprises municipalities where the \textit{Bloque Central Bolívar} continued its expansion during 2000 to 2001. As the system moves toward a denser center with shared intermediation, the strength of the leading actor largely determines internal composition, while the distribution of secondary elements shapes the degree of mixture. The second configuration, which accounts for roughly half of the detected communities, comprises territories where multiple expansive actors operate at scale, but no single one achieves concentrated dominance. These are the most contested localities, where recurrent co-presence sustains high internal interaction with a more even balance between FARC and AUC forces.

The eighth period marks the peak of both rebel and paramilitary expansion. Afterward, the conditions sustaining their growth begin to change. By 2002, the government dismantled the demilitarized zone created for peace talks with the FARC, while the alliances binding the AUC began to fragment. During the ninth period, violence against civilians by FARC and AUC affiliated units reached its highest levels. As the state implemented \textit{Plan Colombia} and intensified operations to dislodge FARC units from \textit{Cundinamarca} to prevent attacks against the capital, intra paramilitary conflict escalated. The \textit{Bloque Central Bolívar}, BCB, declared autonomy within the AUC and, backed by factions aligned with drug trafficking interests, launched a campaign against the \textit{Bloque Metro}, whose leadership opposed financing the war through narcotics.

\begin{figure}[!htb]
    \centering
    \subfigure[Community interaction intensities, $\lambda_{ij}$.]{
        \includegraphics[scale=0.33]{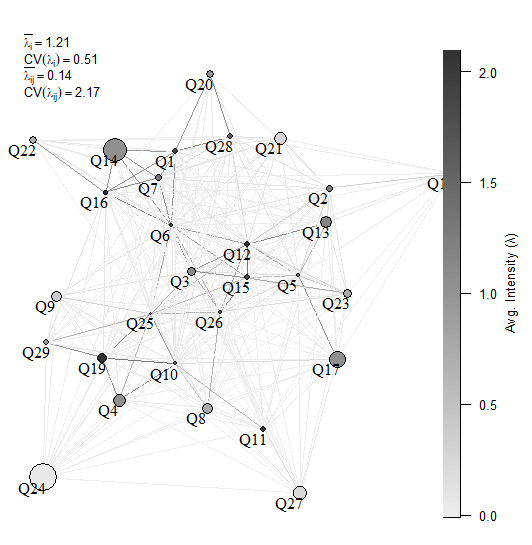}
    }
    \subfigure[Geographical distribution of communities.]{
        \includegraphics[scale=0.48]{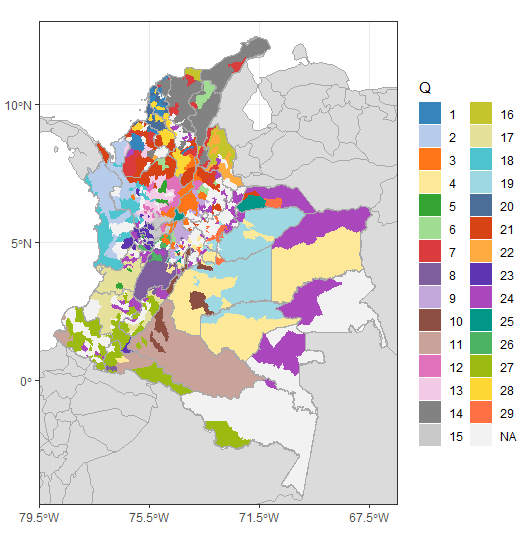}
    }
    \subfigure[Municipality network.]{
        \includegraphics[scale=0.33]{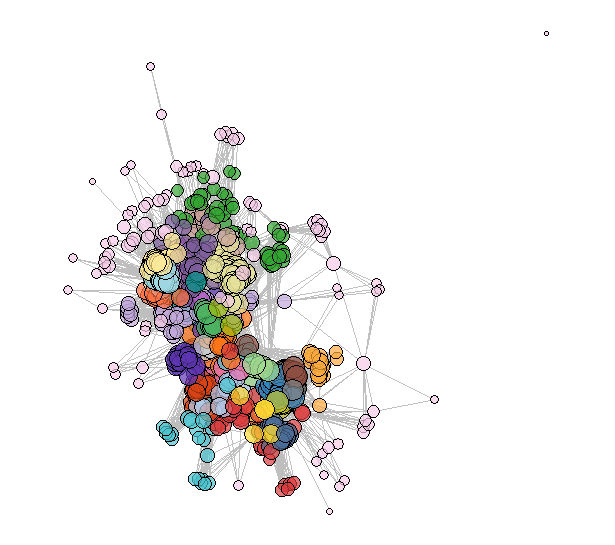}
    }
    \subfigure[$Q_{25}$ ego graph, first-order structure.]{
        \includegraphics[scale=0.33]{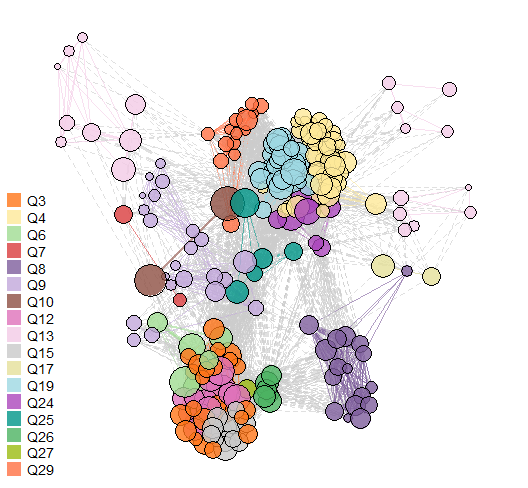}
    }
    \caption{Communities in the municipality network, 2002--2004.}
    \label{fig:cluster_interactions_IX}
\end{figure}

In line with this reorganization of armed structures, the number of communities increases and the size distribution shifts toward smaller and mid sized blocks. Whereas 2000 to 2001 included only one class with fewer than 10 municipalities and a long upper tail up to 67, 2002 to 2004 adds several small sets between 5 and 10 and a thicker middle between 14 and 41, while still featuring large blocks of 81 and 94 municipalities. These results indicate finer differentiation at the center, with satellites that remain attached but are more clearly separated. The SBM predicted within class intensity increases, \(\bar{\lambda}_i\) from 1.08 to 1.21, while the between class mean, \(\bar{\lambda}_{ij}\), shows no meaningful change.

We observe a generalized increase in the area of influence of armed groups, along with greater overlap between them. First, the distribution of municipalities per actor shifts rightward. The median rises from 7 to 9, and the mean from 11.26 to 12.55, indicating broader geographic reach for individual organizations. Second, co-presence at the dyad level thickens. The edge weight distribution extends from a maximum of 5 to 6, and the share of municipality pairs linked by four or more armed structures increases from 0.046\% to 0.301\%, about a 6.6 fold rise.

During this period, the network forms a highly integrated giant component. We observe only three biconnected components, with the largest containing 99\% of vertices, 769 of 772. Thus, almost the entire network lies within a single 2 connected block. Coreness is also very high, with first quartile 19, median 46, and maximum 127. Nearly all municipalities in the large component fall in the 3 core or higher, 766 of about 771, and many have coreness above 50, 329 vertices, and even above 80, 128 vertices. Both the size and internal depth of the giant component are extreme. The network is not only almost fully connected, it is held together by redundant ties.

These shifts capture a denser mesh of shared operating zones. As overlap intensifies and spreads, each location incorporates new local actors absent elsewhere. This pattern reflects the fragmented and fluid character of civil wars, as emphasized by \cite[p.~XX]{Kalyvas2006}. Fragmentation is visible in Figure~\ref{fig:cluster_interactions_IX}, which maps clusters of municipalities. A patchwork structure-emerges, and as disputes intensify, boundaries between opposing sides become less distinct.

Evidence also suggests a redistribution of dominance within communities between 2002 and 2004. Top actor coverage shifts upward, indicating that in more communities at least one group spans a larger share of municipalities within the class. The lower tail rises, from 0.0896 to 0.1064, the first quartile increases from 0.498 to 0.667, and the median remains high at 0.933. At the same time, incidence concentration eases at the center, with median \(\mathrm{HHI}\) moving from 0.194 to 0.168, and evenness becomes less left skewed. The lower tail of \(J_{\mathrm{even}}\) rises, from 0.309 to 0.560, and the mean increases from 0.786 to 0.798. Together, these shifts point to more co anchored compositions. Armed structures cover large shares of territory, but the remaining presence is spread more broadly across additional actors than in 2000 to 2001. Very high coverage is common, yet it more often reflects shared anchoring than monopoly like dominance.

In the low coverage band, \(\mathrm{cov} < 0.4\), incidence concentration remains minimal and evenness stays high, with a slightly higher lower tail, reinforcing a configuration in which no single actor structures within class ties. The band \(0.4 \le \mathrm{cov} < 0.5\) disappears, a redistribution that increases the share of genuinely mixed classes and mid dominance communities. In the mid dominance band, \(0.5 \le \mathrm{cov} < 0.8\), concentration decreases and evenness increases, with medians \(\mathrm{HHI}\) from 0.240 to 0.168 and \(J_{\mathrm{even}}\) from 0.719 to 0.761. This is consistent with two or three co leaders holding comparable shares, without any single actor approaching full territorial reach, and with meaningful residual actors remaining active. In the very high coverage range, \(\mathrm{cov} \ge 0.8\), the medians change little, with \(\mathrm{HHI}\) from 0.168 to 0.182 and \(J_{\mathrm{even}}\) from 0.812 to 0.786, but the extremes move. The upper tail of concentration contracts, with max \(\mathrm{HHI}\) from 0.773 to 0.620, while the lower tail of evenness rises sharply, with min \(J_{\mathrm{even}}\) from 0.309 to 0.560.

Consistent with these results, the reduction in the maximum number of SBM communities connected by between class edges induced by any single dominant actor, from 6 to 5, indicates fewer near monopoly spans and a shift toward near saturation under shared leadership. At the same time, dual dominance becomes more common. The number of distinct actors that significantly influence exactly two communities rises from 15 to 23, driven mainly by paramilitary reconfiguration after the AUC split. As the BCB asserted autonomy in the ninth period, the expansive substructure that shaped connectivity across six municipality communities in 2000 to 2001 reorganized into blocs and fronts, inducing strong ties for specific SBM pairs, northwestern \textit{Antioquia} \(Q_7\) and \(Q_{21}\), northeastern \textit{Antioquia} \(Q_5\) and \(Q_{12}\), and the \textit{Suroeste antioqueño} \(Q_{15}\) and \(Q_{23}\). On the rebel side, civilian targeting rebalanced geographically. Fronts with the broadest reach in 2000 to 2001 were concentrated in the south, \textit{Nariño}, \textit{Putumayo}, and \textit{Caquetá}, whereas in 2002 to 2004 more dominant FARC structures concentrate in the \textit{Montes de María} \(Q_{20}\) and \(Q_{28}\), \(Q_{28}\) and \(Q_7\), the \textit{Urabá} and \textit{Atrato} hinge \(Q_2\) and \(Q_{18}\), the \textit{Oriente antioqueño} \(Q_{12}\) and \(Q_{15}\), and the \textit{Suroeste antioqueño} and \textit{Eje Cafetero} corridor \(Q_{15}\) and \(Q_{23}\).

Additionally, between class connectivity widens. SBM communities relate to more counterpart classes overall, as partner counts increase across the distribution. The median rises from 5.89 to 6.52, the upper quartile from 7.18 to 8.23, and the maximum from 11.56 to 15.18. Heterogeneity also increases because partner counts stretch at both tails. The minimum decreases from 2.745 to 2.272, while the maximum increases from 11.56 to 15.18, and the interquartile range grows from \(7.182 - 4.649 \approx 2.53\) to \(8.231 - 4.369 \approx 3.86\). Some blocks therefore spread contact much more widely, while others remain narrowly connected. At the same time, most outward flow is still concentrated on two counterpart classes. The median share carried by the two strongest counterparts increases slightly, from about 0.594 to 0.602, but the range widens. In 2002 to 2004 it spans 0.296 to 0.974, versus 0.372 to 0.814 in 2000 to 2001. Overall, communities connect to more counterparts on average, yet outward flow either disperses across many partners or is captured almost entirely by two.

\begin{figure}[!htb]
    \centering
    \subfigure[Community interaction intensities, $\lambda_{ij}$.]{
        \includegraphics[scale=0.4]{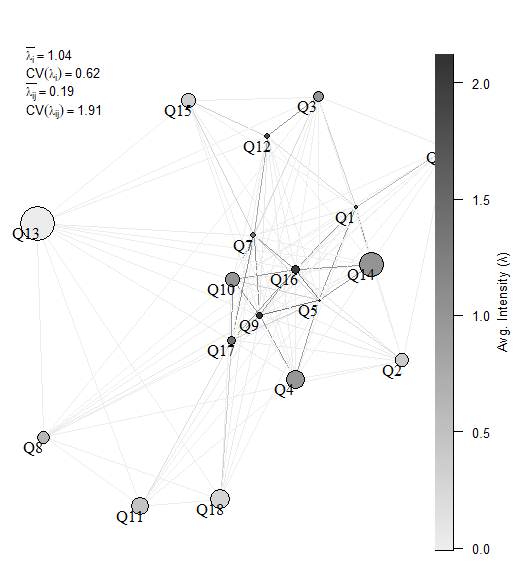}
    }
    \subfigure[Geographical distribution of communities.]{
        \includegraphics[scale=0.4]{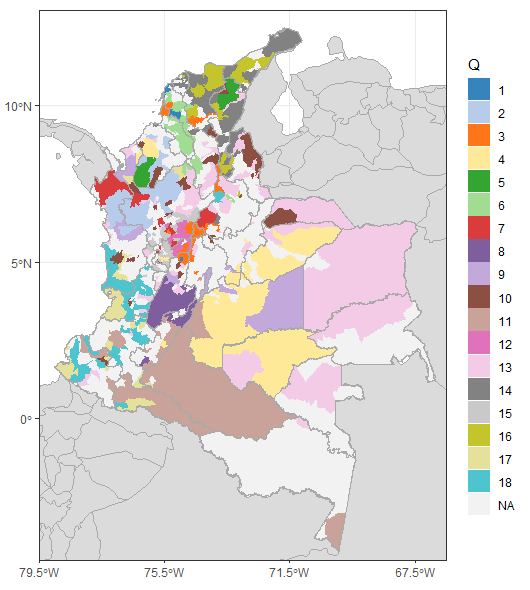}
    }
    \subfigure[Municipality network.]{
        \includegraphics[scale=0.4]{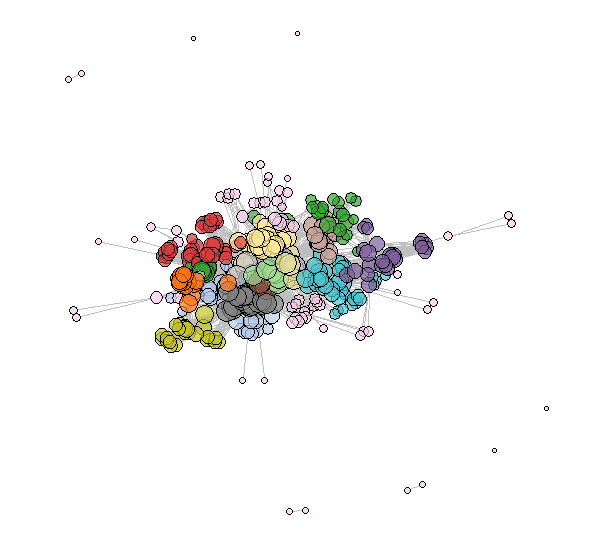}
    }
    \subfigure[$Q_{25}$ ego graph, first-order structure.]{
        \includegraphics[scale=0.4]{plots/core_02_04.png}
    }
    \caption{Communities in the municipality network, 2005--2007.}
    \label{fig:cluster_interactions_X}
\end{figure}

Between 2005 and 2007, the paramilitary project entered a formal demobilization phase. Under the Justice and Peace framework, Law 975 of 2005, the main AUC blocs carried out collective demobilizations, but mid level commanders in key corridors preserved their networks, concealed arsenals, or rearmed soon afterward under new labels. Territories associated with the \textit{Bloque Norte}, \textit{Bloque Centauros}, \textit{Bloque Élmer Cárdenas}, and \textit{Bloque Calima} became focal points of this recycling, as commanders linked to drug trafficking circuits reorganized local units into successor groups such as \textit{Los Urabeños} and \textit{Los Rastrojos}. In parallel, the government deepened counterinsurgency operations against the FARC, combining \textit{Plan Colombia} resources with large-scale offensives that displaced several fronts from \textit{Cundinamarca}, \textit{Meta}, and the cordilleras, but did not eliminate their presence in Pacific and border regions. As a result, the 2005 to 2007 network reflects not only a decline in affected municipalities, but also the contraction of paramilitary corridors into a smaller set of successor strongholds and a more fragmented guerrilla field under sustained military pressure.

The network shows clear thinning and internal segmentation. The number of municipalities with civilian targeting decreases from 772 to 468. The number of biconnected components increases from three to fifteen, indicating that the giant component is now held together by more articulation points, eight more than in the ninth period, and by smaller 2 connected blocks. Coreness also declines sharply, with the median dropping from 46 to 16, the maximum falling from 127 to 81, and the mean from about 53 to 35. Most municipalities in the giant component remain in the 3 core or higher, 435 of roughly 458, so the graph remains genuinely interconnected rather than collapsing into a structure sustained by single paths. Concentration in very high cores weakens. The number of municipalities with coreness above 50, each within a subgraph where every municipality has at least 50 links to others in that same subgraph, falls from 329 to 140, even though the share with coreness above 80 remains similar, about 17\% in both periods.

At the core of the 2005 to 2007 network are the armed structures that re-emerged after the AUC decommissioned its weapons. The SBM clusters the municipalities where these actors are most active into five communities, \(Q_9\), \(Q_{10}\), \(Q_{14}\), \(Q_{16}\), and \(Q_{17}\). We treat this set as the core because it concentrates a large share of predicted between class weight, 44\%, and includes municipalities hosting the most expansive actors during the period. Four clusters fall in territories where former AUC commanders who abandoned demobilization assembled the successor structure known as \textit{Los Urabeños} in their previous strongholds, and the fifth captures its intersection with the remaining units of the \textit{Bloque Norte}.

Each core cluster maps a distinct configuration of territorial overlap. Communities \(Q_9\) and \(Q_{17}\) both reflect the origins of \textit{Los Urabeños}, but they differ in geography and in their surrounding actors. In \(Q_9\), the network captures a tight linkage between the post-demobilization group and the \textit{Bloque Centauros}. This community is not geographically compact, but spans municipalities in \textit{Urabá} and the \textit{Llanos Orientales}, highlighting the persistence of ties between these regions after paramilitary disarmament. Consistent with intense contestation, municipality pairs in \(Q_9\) are typically linked by more than two armed structures, the highest overlap in the system during this period. By contrast, \(Q_{17}\) gathers municipalities where \textit{Los Urabeños} co-occur with actors concentrated in the southwestern Pacific, including FARC fronts and \textit{Los Rastrojos}, so co-presence remains locally bounded rather than linking distant regions. Community \(Q_{10}\) groups 30 municipalities distributed across 25 subregions in 14 departments, a set of secondary strongholds where \textit{Los Urabeños} appear without other armed units targeting civilians, pointing to expansion into former AUC territories with weaker competition. Community \(Q_{16}\) concentrates municipalities in \textit{Magdalena}, \textit{La Guajira}, and \textit{Cesar}, where \textit{Los Urabeños} and the \textit{Bloque Norte} still co-occur, while \(Q_{14}\) isolates nearby municipalities in the same departments where the latter operates alone. Together, these two classes delineate the residual Caribbean corridor of the AUC and its realignment alongside the emerging successor group.

The periphery comprises thirteen classes and about 73\% of municipalities, 345 of 468. Despite this broad territorial reach, ties among peripheral blocs account for only about 18\% of total between class weight, while nearly 39\% runs along core periphery edges. Peripheral communities combine lower within class degree, \(\bar d_{\text{per}} \approx 11.8\) versus \(\bar d_{\text{core}} \approx 32.9\), with higher dispersion in local connectivity, with mean \(cv^{\text{intra}}_{\text{per}} \approx 43\) and \(cv^{\text{intra}}_{\text{core}} \approx 0\). Clustering is also lower in eight of these blocs than in the core, which is formed by complete cliques, suggesting more uneven and localized structures.

These areas reflect a shift in the rebel paramilitary balance away from the post AUC core. They form an outer belt in which FARC influence is comparatively stronger and between class connectivity is more fragmented. Several are predominantly guerrilla based, including \(Q_2\), \(Q_8\), and \(Q_{11}\), where 60\% to 90\% of active substructures are FARC units and paramilitary or post conflict groups appear only marginally. Others, such as \(Q_3\), \(Q_5\), \(Q_6\), \(Q_{13}\), \(Q_{15}\), and \(Q_{18}\), are mixed zones where rebel and right wing structures coexist in varying proportions, often with smaller post-demobilization or organized crime actors present as well.

\newpage

Within this weak periphery periphery connectivity, a secondary hub emerges around \(Q_7\), \(Q_{12}\), and \(Q_{15}\), adjacent to the paramilitary core. Community \(Q_7\) covers municipalities in eastern \textit{Antioquia} where FARC fronts attached to the \textit{Bloque Noroccidental} contest territorial control with the successor paramilitary formation that replaced the AUC. Communities \(Q_{12}\) and \(Q_{15}\) extend this interface into adjacent municipalities where the same FARC fronts intersect remnant AUC units such as the \textit{Bloque Mineros}, \textit{Bloque Héroes de Granada}, and \textit{Bloque Central Bolívar}, together with smaller post-demobilization structures.

We therefore conclude that when a subset of armed structures dominates a territory and bilateral interactions intensify, the network tends toward a node periphery structure with high centralization, as in 1985 to 1987 and 1994 to 1996. Conversely, when territories become more differentiated, the network resembles a union of cliques in which a relatively small set of nodes concentrates intermediary positions, bridging sectors that remain weakly connected, as in 1997 to 1999. These nodes likely correspond to areas of greatest contention. In contrast, when the number of territories shared by armed structures increases, cliques overlap more and connection strength rises more broadly, not only in the upper tail, indicating a more diffuse and widespread territorial dispute, as in 2000 to 2004.

\section{Discussion}

The topological analysis of the municipality networks shows that the system’s structure changes substantially across periods. The organizational transformations of armed groups through expansion, consolidation, disarmament, and rearmament between 1978 and 2007 are reflected in the networks’ structural features. When a limited set of armed structures achieves broad territorial dominance, the network tends to a centralized node periphery configuration with high centralization. This pattern is most evident in 1985 to 1987 and 1994 to 1996, when bilateral interactions concentrate in a few central nodes that connect a large share of the network. During phases of greater territorial differentiation, such as 1997 to 1999, the network resembles a union of cliques, where a small number of nodes mediate between otherwise weakly connected regions, suggesting areas of intense dispute. Finally, when armed structures share a larger number of territories, as in 2000 to 2004, link strength increases more broadly across the network, indicating that territorial disputes extend to a wider set of regions. Overall, these findings show how the fragmentation and reconfiguration of territorial control reshape interaction patterns as civil war dynamics evolve.

The stochastic block model fitted to the municipality network identifies groups of municipalities using the observed ties and their strengths. The results suggest communities shaped not only by geographic proximity, but also by affinities among armed structures and their leaderships. At the same time, we identify groups with high internal dispersion, indicating communities with very few internal connections, while in other cases between community interaction is comparatively high. This pattern suggests that during periods of intense territorial dispute, the conflict does not yield clear separations between opposing substructures, unlike interstate wars characterized by stable front lines. These results are consistent with Kalyvas’s argument that territorial control in civil wars is fragmented and fluid.

The model reproduces the network’s first order structure, particularly the mean vertex strength and its dispersion, but it has clear limitations for density, transitivity, and maximum coreness. Posterior predictive \(p\) values near zero indicate systematic underestimation of overall connectivity, triadic closure, and core depth. In practice, the fitted model rarely generates networks that are as dense, as clustered, or with cores as deep as those observed, so it does not fully capture the concentration of ties around a subset of highly embedded municipalities. This limitation matters in civil war settings, where it is crucial to determine whether violence against civilians is organized around a few territorial focal points or is more diffusely contested. Future work could generalize the specification to better represent higher order concentration, for example by introducing mechanisms that directly affect clustering or core depth.

We restrict the analysis to two factions because the variables used to attribute responsibility to specific armed structures are of low quality. Extending the study to additional actors, including other guerrillas such as ELN, EPL, and M 19, and state forces, requires substantial data cleaning. We expect to incorporate these actors in future work.

Missing data are also substantial, particularly for enforced disappearances and sexual violence. This occurs partly because victims may not know the identity of the perpetrator, and partly because of the difficulties of documentation and institutional constraints in managing records. Missingness can bias the characterization of the phenomenon, so addressing it is an important direction for future research.

This research contributes a methodological and analytical framework that reveals dimensions of internal armed conflict that are often missed by purely qualitative or chronological approaches. By integrating historical records from the \textit{Centro Nacional de Memoria Histórica} with network analysis, the study identifies not only who the actors are and where they operate, but also how their relationships and territorial configurations change over time. This dynamic perspective uncovers structural patterns, including fragmentation, consolidation, and overlapping spheres of influence, that help explain the persistence and transformation of violence. Rather than treating conflict as a linear sequence of events, the approach models it as an evolving system in which the configuration of relationships between armed actors and territories shapes the conflict’s trajectory.

The approach also extends beyond the Colombian case. It can be applied to conflicts in other regions, used for cross country comparisons, or adapted to other periods of Colombia’s history. The framework can incorporate additional layers, such as economic networks, political alliances, illicit economies, or displacement patterns, to enrich the characterization of the broader wartime system. By documenting how armed groups form, dissolve, and reorganize, this study challenges static conceptions of conflict and offers a replicable strategy for analyzing the fluidity of violence in complex settings. These insights can inform academic debates and support policymakers, peace negotiators, and transitional justice practitioners seeking interventions that target the structural and relational conditions that allow armed violence to persist or re-emerge.

Finally, this investigation does not seek to reduce victims of the Colombian armed conflict, or the gravity of the crimes committed, to numbers or statistical outputs. Quantitative methods are used as tools to study the conflict’s structure and dynamics, identifying patterns and transformations that are difficult to detect through isolated accounts. We recognize that each data point represents a person, a family, and a community that has endured profound suffering and irreparable loss. This recognition is central to the study. Statistical models and network visualizations are instruments, not ends. The goal is to provide rigorous evidence that can inform public policy, guide institutional decisions, and support government actors and transitional justice mechanisms in designing interventions aimed at ensuring that such crimes are not repeated.

\section*{Statements and declarations}

The data supporting this publication can be obtained from the website of the \textit{Centro Nacional de Memoria Histórica}. They are publicly accessible under the terms and conditions set forth by the source and can be accessed at the following link: \url{https://micrositios.centrodememoriahistorica.gov.co/}.

The authors declare that they have no known competing financial interests or personal relationships that could have appeared to influence the work reported in this article.

During the preparation of this work the authors used ChatGPT-5-Thinking in order to improve language and readability. After using this tool, the authors reviewed and edited the content as needed and take full responsibility for the content of the publication.

\bibliography{references.bib}

@article{clauset2009,
  author  = {Clauset, A. and Shalizi, C. R. and Newman, M. E. J.},
  title   = {Power-law distributions in empirical data},
  journal = {SIAM Review},
  volume  = {51},
  number  = {4},
  pages   = {661--703},
  year    = {2009},
  doi     = {10.1137/070710111},
  url     = {https://doi.org/10.1137/070710111}
}

@article{Gillespie2015,
  author    = {Colin S. Gillespie},
  title     = {Fitting Heavy Tailed Distributions: The powerlaw Package},
  journal   = {Journal of Statistical Software},
  year      = {2015},
  volume    = {64},
  number    = {2},
  pages     = {1--16},
  url       = {http://www.jstatsoft.org/v64/i02/}
}

@book{MedinaGallego2011,
  author    = {Medina Gallego, C. and others},
  title     = {FARC-EP: flujos y reflujos. La guerra en las regiones},
  year      = {2011},
  address   = {Bogotá},
  publisher = {Universidad Nacional de Colombia, Facultad de Derecho, Ciencias Políticas y Sociales, Instituto Unidad de Investigaciones Jurídico-Sociales Gerardo Molina (UNIJUS)},
  pages     = {324},
  note      = {Incluye referencias bibliográficas},
  isbn      = {978-958-719-927-7},
  langid    = {spanish}
}

@book{CNMH2017b,
  author    = {CNMH},
  title     = {Medellín, memorias de una guerra urbana},
  year      = {2017},
  address   = {Bogotá},
  publisher = {Centro Nacional de Memoria Histórica, Corporación Región, Ministerio del Interior, Alcaldía de Medellín, Universidad EAFIT, Universidad de Antioquia},
  series    = {¡Basta ya! Regionales},
  pages     = {524},
  note      = {Tablas, gráficos, mapas, fotografías},
  isbn      = {978-958-8944-73-9},
  langid    = {spanish}
}

@book{CEV2022a,
  author    = {CEV},
  title     = {Colombia Adentro. Relatos territoriales sobre el conflicto armado. Nariño y Sur de Cauca},
  year      = {2022},
  edition   = {1},
  address   = {Bogotá},
  publisher = {Comisión para el Esclarecimiento de la Verdad, la Convivencia y la No Repetición},
  note      = {Tomo 11, volumen 8 del \textit{Informe Final Hay futuro si hay verdad}},
  isbn      = {978-628-7590-35-9},
  langid    = {spanish},
  howpublished = {\url{https://www.comisiondelaverdad.co/}}
}

@book{CEV2022b,
  author    = {CEV},
  title     = {Colombia Adentro. Relatos territoriales sobre el conflicto armado. Amazonía},
  year      = {2022},
  edition   = {1},
  address   = {Bogotá},
  publisher = {Comisión para el Esclarecimiento de la Verdad, la Convivencia y la No Repetición},
  note      = {Tomo 11, volumen 2 del \textit{Informe Final Hay futuro si hay verdad}},
  isbn      = {978-628-7590-30-4},
  langid    = {spanish},
  howpublished = {\url{https://www.comisiondelaverdad.co/}}
}

@book{CEV2022c,
  author    = {CEV},
  title     = {Colombia Adentro. Relatos territoriales sobre el conflicto armado. Orinoquía},
  year      = {2022},
  edition   = {1},
  address   = {Bogotá},
  publisher = {Comisión para el Esclarecimiento de la Verdad, la Convivencia y la No Repetición},
  note      = {Tomo 11, volumen 9 del \textit{Informe Final Hay futuro si hay verdad}},
  isbn      = {978-628-7590-36-6},
  langid    = {spanish},
  howpublished = {\url{https://www.comisiondelaverdad.co/}}
}

@book{CEV2022d,
  author    = {CEV},
  title     = {Colombia Adentro. Relatos territoriales sobre el conflicto armado. Antioquia},
  year      = {2022},
  edition   = {1},
  address   = {Bogotá},
  publisher = {Comisión para el Esclarecimiento de la Verdad, la Convivencia y la No Repetición},
  note      = {Tomo 11, volumen 3 del \textit{Informe Final Hay futuro si hay verdad}},
  isbn      = {978-628-7590-31-1},
  langid    = {spanish},
  howpublished = {\url{https://www.comisiondelaverdad.co/}}
}

@book{CEV2022e,
  author    = {CEV},
  title     = {Colombia Adentro. Relatos territoriales sobre el conflicto armado. Región Centro},
  year      = {2022},
  edition   = {1},
  address   = {Bogotá},
  publisher = {Comisión para el Esclarecimiento de la Verdad, la Convivencia y la No Repetición},
  note      = {Tomo 11, volumen 11 del \textit{Informe Final Hay futuro si hay verdad}},
  isbn      = {978-628-7590-42-7},
  langid    = {spanish},
  howpublished = {\url{https://www.comisiondelaverdad.co/}}
}

@book{CEV2022f,
  author    = {CEV},
  title     = {Colombia Adentro. Relatos territoriales sobre el conflicto armado. Magdalena Medio},
  year      = {2022},
  edition   = {1},
  address   = {Bogotá},
  publisher = {Comisión para el Esclarecimiento de la Verdad, la Convivencia y la No Repetición},
  note      = {Tomo 11, volumen 7 del \textit{Informe Final Hay futuro si hay verdad}},
  isbn      = {978-628-7590-34-2},
  langid    = {spanish},
  howpublished = {\url{https://www.comisiondelaverdad.co/}}
}

@book{CEV2022g,
  author    = {CEV},
  title     = {Colombia Adentro. Relatos territoriales sobre el conflicto armado. Pacífico},
  year      = {2022},
  edition   = {1},
  address   = {Bogotá},
  publisher = {Comisión para el Esclarecimiento de la Verdad, la Convivencia y la No Repetición},
  note      = {Tomo 11, volumen 10 del \textit{Informe Final Hay futuro si hay verdad}},
  isbn      = {978-628-7590-37-3},
  langid    = {spanish},
  howpublished = {\url{https://www.comisiondelaverdad.co/}}
}

@book{CEV2022h,
  author    = {CEV},
  title     = {Colombia Adentro. Relatos territoriales sobre el conflicto armado. Valle y Norte del Cauca},
  year      = {2022},
  edition   = {1},
  address   = {Bogotá},
  publisher = {Comisión para el Esclarecimiento de la Verdad, la Convivencia y la No Repetición},
  note      = {Tomo 11, volumen 12 del \textit{Informe Final Hay futuro si hay verdad}},
  isbn      = {978-628-7590-38-0},
  langid    = {spanish},
  howpublished = {\url{https://www.comisiondelaverdad.co/}}
}

@book{CEV2022j,
  author    = {CEV},
  title     = {Colombia Adentro. Relatos territoriales sobre el conflicto armado. Caribe},
  year      = {2022},
  edition   = {1},
  address   = {Bogotá},
  publisher = {Comisión para el Esclarecimiento de la Verdad, la Convivencia y la No Repetición},
  note      = {Tomo 11, volumen 8 del \textit{Informe Final Hay futuro si hay verdad}},
  isbn      = {978-628-7590-41-0},
  langid    = {spanish},
  howpublished = {\url{https://www.comisiondelaverdad.co/}}
}

@article{Velez2001,
  author  = {Vélez, M},
  title   = {FARC-ELN: evolución y expansión territorial},
  journal = {Desarrollo y Sociedad},
  year    = {2001},
  volume  = {47},
  month   = {marzo},
  pages   = {},
  langid  = {spanish},
  note    = {Published by Universidad de los Andes, Department of Economics}
}

@article{Biernacki2000,
  author = {Biernacki, Christophe and Celeux, Gilles and Govaert, Georges},
  title = {Assessing a Mixture Model for Clustering with the Integrated Completed Likelihood},
  journal = {IEEE Transactions on Pattern Analysis and Machine Intelligence},
  volume = {22},
  number = {7},
  pages = {719-725},
  year = {2000},
  doi = {10.1109/34.846210}
}

@manual{R2019,
  title = {R: A Language and Environment for Statistical Computing},
  author = {{R Core Team}},
  year = {2019},
  organization = {R Foundation for Statistical Computing},
  address = {Vienna, Austria},
  url = {https://www.R-project.org/}
}

@article{Borgatti2016,
  author = {Borgatti, S. P. and Carley, K. M. and Krackhardt, D.},
  title = {Blockmodels: A R-package for estimating in Latent Block Model and Stochastic Block Model, with various probability functions, with or without covariates},
  journal = {Journal of Statistical Software},
  volume = {70},
  number = {1},
  pages = {1-24},
  year = {2016},
  url = {https://doi.org/10.18637/jss.v070.i01}
}

@book{Arjona2016,
  author    = {Ana Arjona},
  title     = {Rebelocracy: Social Order in the Colombian Civil War},
  year      = {2016},
  publisher = {Cambridge University Press},
  address   = {Cambridge},
}

@book{bondy1976graph,
  title={Graph Theory with Applications},
  author={Bondy, John Adrian and Murty, U. S. R.},
  year={1976},
  publisher={Elsevier}
}

@book{diestel2017graph,
  title={Graph Theory},
  author={Diestel, Reinhard},
  year={2017},
  edition={5th},
  publisher={Springer}
}

@book{west2001introduction,
  title={Introduction to Graph Theory},
  author={West, Douglas B.},
  year={2001},
  edition={2nd},
  publisher={Prentice Hall}
}

@article{holland1983stochastic,
  title={Stochastic blockmodels: First steps},
  author={Holland, Paul W. and Laskey, Kathryn Blackmond and Leinhardt, Samuel},
  journal={Social Networks},
  volume={5},
  number={2},
  pages={109--137},
  year={1983},
  publisher={Elsevier}
}

@article{nowicki2001estimation,
  title={Estimation and prediction for stochastic blockstructures},
  author={Nowicki, Krzysztof and Snijders, Tom AB},
  journal={Journal of the American Statistical Association},
  volume={96},
  number={455},
  pages={1077--1087},
  year={2001},
  publisher={Taylor \& Francis}
}

@article{aicher2014learning,
  title={Learning latent block structure in weighted networks},
  author={Aicher, Christopher and Jacobs, Abigail Z and Clauset, Aaron},
  journal={Journal of Complex Networks},
  volume={3},
  number={2},
  pages={221--248},
  year={2014},
  publisher={Oxford University Press}
}

@book{CNMH2013,
  title={Guerrilla y población civil: trayectoria de las FARC 1949-2013},
  author={CNMH},
  year={2013},
  publisher={CNMH},
  isbn={978-958-58167-0-1}
}

@book{CNMH2017,
  title={De los grupos precursores al bloque Tolima AUC. Informe N.o 1, Serie: Informes sobre el origen y actuación de las agrupaciones paramilitares en las regiones, Bogotá D.C},
  author={CNMH},
  year={2017},
  publisher={CNMH},
  isbn={978-958-8944-76-0}
}

@book{CNMH2018a,
  title={Bloque Calima de las AUC: depredación paramilitar y narcotráfico en el suroccidente colombiano. Informe N.o 2, Serie: Informes sobre el origen y actuación de las agrupaciones paramilitares en las regiones, Bogotá D.C},
  author={CNMH},
  year={2018},
  publisher={CNMH},
  isbn={978-958-8944-90-6}
}

@book{CNMH2018b,
  title={Violencia paramilitar en la Altillanura: autodefensas campesinas de Meta y Vichada. Informe N.o 3, Serie: Informes sobre el origen y actuación de las agrupaciones paramilitares en las regiones, Bogotá D.C},
  author={CNMH},
  year={2018},
  publisher={CNMH},
  isbn={978-958-5500-39-6}
}

@book{CNMH2019a,
  title={El estado suplantado: las autodefensas de Puerto Boyacá. Informe N.o 4, Serie: Informes sobre el origen y actuación de las agrupaciones paramilitares en las regiones},
  author={CNMH},
  year={2019},
  publisher={CNMH},
  isbn={978-958-5500-52-5}
}

@book{CNMH2019b,
  title={El modelo paramilitar de San Juan Bosco de La Verde y Chucurí. Informe N.o 5, Serie: Informes sobre el origen y la actuación de las agrupaciones paramilitares en las regiones},
  author={CNMH},
  year={2019},
  publisher={CNMH},
  isbn={978-958-5500-54-9}
}

@book{CNMH2020a,
  title={Isaza, el clan paramilitar: Las Autodefensas Campesinas del Magdalena Medio. Informe N.o 6, Serie: Informes sobre el origen y la actuación de las agrupaciones paramilitares en las regiones},
  author={CNMH},
  year={2020},
  publisher={CNMH},
  isbn={978-958-5500-55-6}
}

@book{CNMH2020b,
  title={Autodefensas de Cundinamarca: olvido estatal y violencia paramilitar en las provincias de Rionegro y Bajo Magdalena},
  author={CNMH},
  year={2020},
  publisher={CNMH},
  isbn={978-958-5500-56-3},
  note={ISBN digital: 978-628-7561-05-2}
}

@book{CNMH2021a,
  title={Doble discurso, múltiples crímenes: Análisis temático de las ACMM y las ACPB. Informe N.o 9, Serie: Informes sobre el origen y la actuación de las agrupaciones paramilitares en las regiones},
  author={CNMH},
  year={2021},
  publisher={CNMH},
  isbn={978-958-5500-70-9},
  note={ISBN digital: 978-958-5500-71-6}
}

@book{CNMH2021b,
  title={Memorias de una guerra por los Llanos. Tomo I: De la violencia a las resistencias ante el Bloque Centauros de las AUC},
  author={CNMH},
  year={2021},
  publisher={CNMH},
  isbn={978-958-5500-78-5},
  note={ISBN digital: 978-958-5500-79-2}
}

@book{CNMH2021c,
  title={Memorias de una guerra por los Llanos. Tomo II: El Frente Capital y el declive del Bloque Centauros de las AUC},
  author={CNMH},
  year={2021},
  publisher={CNMH},
  isbn={978-958-5500-84-6},
  note={ISBN digital: 978-958-5500-85-3}
}

@book{CNMH2021d,
  title={Arrasamiento y control paramilitar en el sur de Bolívar y Santander. Tomo I: Bloque Central Bolívar: origen y consolidación},
  author={CNMH},
  year={2021},
  publisher={CNMH},
  isbn={978-958-5500-59-4},
  note={ISBN digital: 978-628-7561-11-3}
}

@book{CNMH2021e,
  title={Arrasamiento y control paramilitar en el sur de Bolívar y Santander. Tomo II: Bloque Central Bolívar: violencia pública y resistencias no violentas},
  author={CNMH},
  year={2021},
  publisher={CNMH},
  isbn={ISBN digital: 978-958-5500-64-8}
}

@book{CNMH2022a,
  title={La tierra se quedó sin su canto: Trayectoria e impactos del Bloque Norte en los departamentos de Atlántico, Cesar, La Guajira y Magdalena. Tomo I. Informe N.o 11, Serie: Informes sobre el origen y la actuación de las agrupaciones paramilitares en las regiones},
  author={CNMH},
  year={2022},
  publisher={CNMH},
  isbn={ISBN digital: 978-958-5500-86-0}
}

@book{CNMH2022b,
  title={La tierra se quedó sin su canto: Trayectoria e impactos del Bloque Norte en los departamentos de Atlántico, Cesar, La Guajira y Magdalena. Tomo II. Informe N.o 11, Serie: Informes sobre el origen y la actuación de las agrupaciones paramilitares en las regiones},
  author={CNMH},
  year={2022},
  publisher={CNMH},
  isbn={ISBN digital:978-958-5500-87-7}
}

@book{CNMH2022c,
  title={Bloque Mineros de las AUC: violencia contrainsurgente, economías criminales y depredación sexual. Informe N.o 12, Serie: Informes sobre el origen y la actuación de las agrupaciones paramilitares en las regiones},
  author={CNMH},
  year={2022},
  publisher={CNMH},
  isbn={ISBN digital: 978-958-5500-96-9}
}

@book{CNMH2022d,
  title={Estrategias de guerra y trasfondos del paramilitarismo en el Urabá antioqueño, sur de Córdoba, bajo Atrato y Darién. Tomo I},
  author={CNMH},
  year={2022},
  publisher={CNMH},
  isbn={ISBN digital: 978-628-7561-18-2}
}

@book{CNMH2022e,
  title={Estrategias de guerra y trasfondos del paramilitarismo en el Urabá antioqueño, sur de Córdoba, bajo Atrato y Darién. Tomo II},
  author={CNMH},
  year={2022},
  publisher={CNMH},
  isbn={ISBN digital: 978-628-7561-21-2}
}

@book{CNMH2022f,
  title={La guerra vino de afuera: el Bloque Pacífico en el sur del Chocó: una herida que aún no cierra. Informe N.o 14, Serie: Informes sobre el origen y la actuación de las agrupaciones paramilitares en las regiones},
  author={CNMH},
  year={2022},
  publisher={CNMH},
  isbn={ISBN digital: 978-628-7561-27-4}
}

@book{CNMH2022g,
  title={Y llegaron por el río: Bloque Vencedores de Arauca 2001–2005. Informe N.o 15, Serie: Informes sobre el origen y la actuación de las agrupaciones paramilitares en las regiones},
  author={CNMH},
  year={2022},
  publisher={CNMH},
  isbn={ISBN digital: 978-628-7561-33-5}
}

@book{CNMH2022h,
  title={El Bloque Central Bolívar y la expansión de la violencia paramilitar. Tomo I. “Mataron a la gente por matarla”: El BCB en Antioquia y el Eje Cafetero. Informe N.o 16, Serie: Informes sobre el origen y la actuación de las agrupaciones paramilitares en las regiones},
  author={CNMH},
  year={2022},
  publisher={CNMH},
  isbn={ISBN digital: 978-628-7561-35-9}
}

@book{CNMH2022j,
  title={Un poco de verdad para poder respirar. Trayectorias e impactos de los bloques paramilitares Montes de María y Mojana. Informe N.° 17, Serie: Informes sobre el origen y la actuación de las agrupaciones paramilitares en las regiones},
  author={CNMH},
  year={2022},
  publisher={CNMH},
  isbn={ISBN digital: 978-628-7561-54-0}
}

@book{CNMH2023a,
  title={El Bloque Central Bolívar y la expansión de la violencia paramilitar. Tomo II. “Todo el mundo sabía que eran ellos”: el BCB en Nariño, Putumayo, Caquetá y los Llanos Orientales. Informe N.o 18, Serie: Informes sobre el origen y la actuación de las agrupaciones paramilitares en las regiones},
  author={CNMH},
  year={2023},
  publisher={CNMH},
  isbn={ISBN digital: 978-628-7561-58-8}
}

@book{CNMH2023b,
  title={Bloque Central Bolívar y la expansión de la violencia paramilitar. Tomo III. Quisieron matar la utopía: la imposición del orden no deseado. Informe N.o 19, Serie: Informes sobre el origen y la actuación de las agrupaciones paramilitares en las regiones},
  author={CNMH},
  year={2023},
  publisher={CNMH},
  isbn={ISBN digital: 978-628-7561-64-9}
}

@book{CNMH2023c,
  title={El estallido de un trueno ajeno. Memorias de sobrevivientes al Bloque Catatumbo. Tomo I. Informe N.o 20, Serie: Informes sobre el origen y la actuación de las agrupaciones paramilitares en las regiones},
  author={CNMH},
  year={2023},
  publisher={CNMH},
  isbn={ISBN digital: 978-628-7561-68-7}
}

@book{CNMH2023d,
  title={Guerra sin fronteras, resistencias sin límites. Memorias de sobrevivientes al Bloque Catatumbo. Tomo II. Informe N.o 20, Serie: Informes sobre el origen y la actuación de las agrupaciones paramilitares en las regiones},
  author={CNMH},
  year={2023},
  publisher={CNMH},
  isbn={ISBN digital: 978-628-7561-70-0}
}

@book{CNMH2023e,
  title={Violencia y estigmatización social en el sur del Cesar y en la provincia de Ocaña},
  author={CNMH},
  year={2023},
  publisher={CNMH},
  isbn={ISBN digital: 978-628-7561-72-4}
}

@book{CNMH2023f,
  title={Dirección de acuerdos de la verdad, YO APORTO A LA VERDAD. Acuerdos de contribución a la verdad y la memoria histórica. Mecanismo no judicial de contribución a la verdad, la memoria histórica y la reparación},
  author={CNMH},
  year={2023},
  publisher={CNMH},
  isbn={ISBN digital: 978-958-8469-86-7}
}

@book{Kalyvas2006,
  title={The Logic of Violence in Civil War},
  author={Kalyvas, Stathis N.},
  year={2006},
  publisher={Cambridge University Press},
  address={Cambridge},
  doi={10.1017/CBO9780511818462}
}

@article{KarrerNewman2010,
  title={Stochastic blockmodels and community structure in networks},
  author={Karrer, Brian and Newman, M. E. J.},
  journal={Physical Review E},
  volume={83},
  number={1},
  pages={016107},
  year={2011},
  publisher={American Physical Society},
  doi={10.1103/PhysRevE.83.016107}
}

@book{KolaczykCsardi2014,
  title={Statistical Analysis of Network Data with R},
  author={Kolaczyk, Eric D. and Csárdi, Gábor},
  year={2014},
  publisher={Springer},
  address={New York},
  doi={10.1007/978-1-4939-0983-4}
}

@book{Luke2015,
  author    = {Douglas A. Luke},
  title     = {A User's Guide to Network Analysis in R},
  year      = {2015},
  publisher = {Springer},
  address   = {Cham, Switzerland},
  series    = {Use R!},
  isbn      = {978-3-319-23882-1},
  doi       = {10.1007/978-3-319-23883-8},
}

@book{menzel2010introduction,
  title     = {Introduction to Social Network Methods},
  author    = {Robert W. Menzel},
  year      = {2010},
  publisher = {University of California, Riverside},
  url       = {http://www.faculty.ucr.edu/~hbarker/socialnetworkmethods/}
}

@misc{RamaJudicial,
  author       = {{Consejo de Estado}},
  title        = {Graves violaciones a los derechos humanos},
  year         = {2019},
  note         = {Accedido: 21 de abril de 2024},
  url          = {https://sidn.ramajudicial.gov.co/SIDN/DOCTRINA/TEXTOS_COMPLETOS/GravesViolaciones/},
}
\bibliographystyle{apalike}

\appendix

\section{Bipartite networks}

Blue nodes represent paramilitary armed substructures, purple nodes represent FARC armed substructures, and orange nodes represent organized crime. Yellow nodes correspond to the municipalities where these substructures were active. Edges represent the links between armed substructures and municipalities, as defined in the previous section. Depending on the projection, edge weights capture either the number of municipalities connected to an armed structure or the number of armed groups associated with a municipality during the period.

\begin{figure}[!htb]
    \centering
    \subfigure[1978--1981.]{
        \includegraphics[scale=0.3]{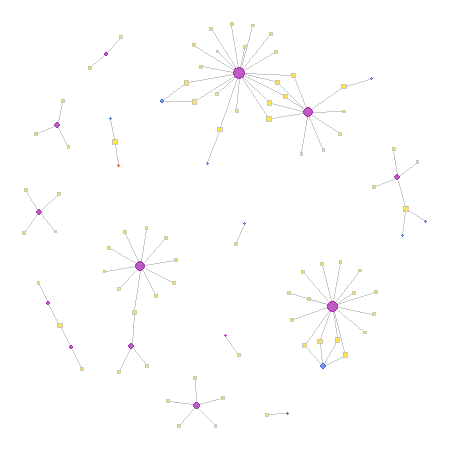}
    }
    \subfigure[1982--1984.]{
        \includegraphics[scale=0.3]{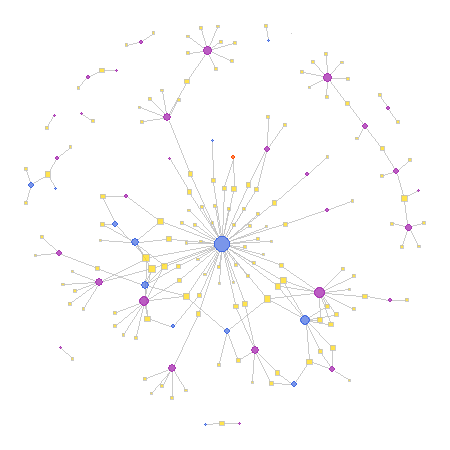}
    }
    \subfigure[1985--1987.]{
        \includegraphics[scale=0.3]{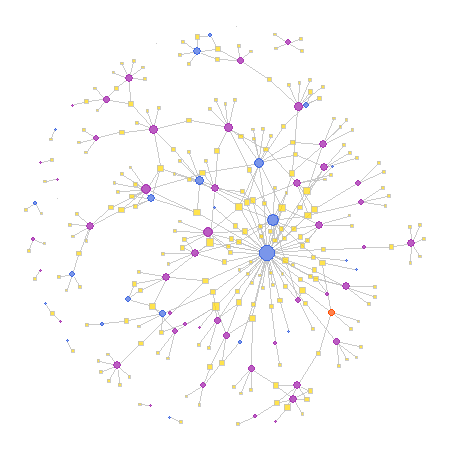}
    }
    \subfigure[1988--1990.]{
        \includegraphics[scale=0.3]{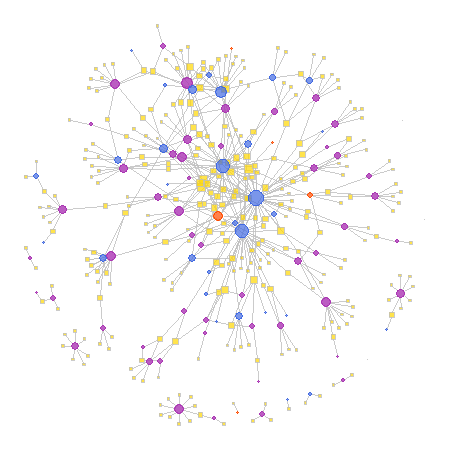}
    }
    \caption{Bipartite networks of armed structures--municipalities.}
    \label{fig:Redes_Bipartitas1}
\end{figure}

\begin{figure}[!htb]
    \centering
    \subfigure[1991--1993.]{
        \includegraphics[scale=0.3]{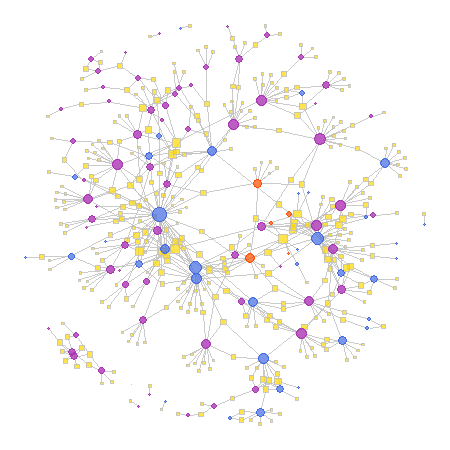}
    }
    \subfigure[1994--1996.]{
        \includegraphics[scale=0.3]{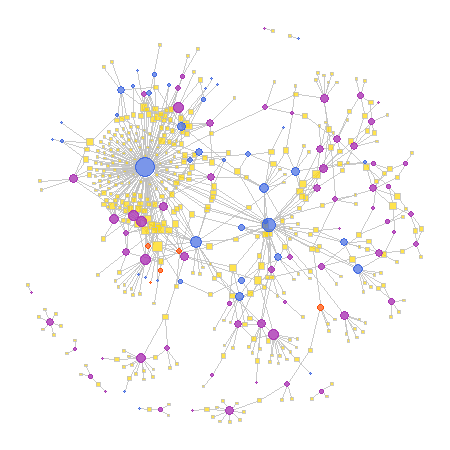}
    }
    \subfigure[1997--1999.]{
        \includegraphics[scale=0.3]{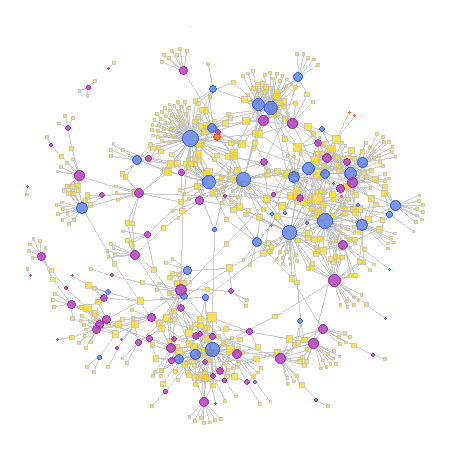}
    }
    \subfigure[2000--2001.]{
        \includegraphics[scale=0.3]{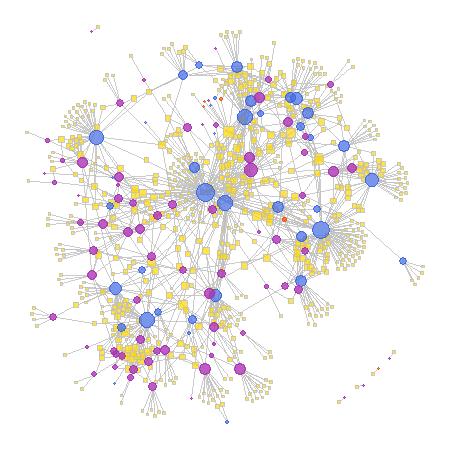}
    }
    \subfigure[2002--2004.]{
        \includegraphics[scale=0.3]{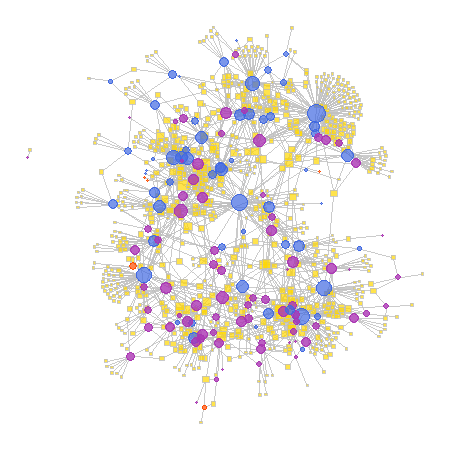}
    }
    \subfigure[2005--2007.]{
        \includegraphics[scale=0.3]{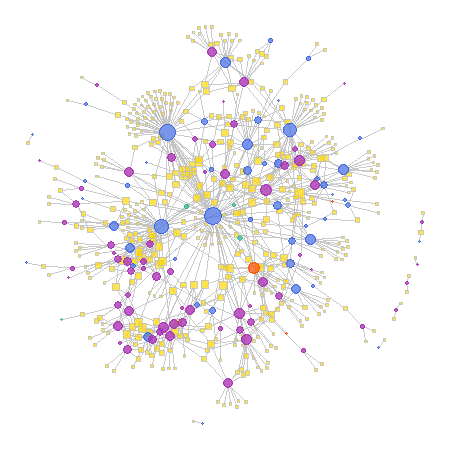}
    }
    \caption{Bipartite networks of municipalities--armed structures.}
    \label{fig:Redes_Bipartitas2}
\end{figure}

\section{Integration Classification Likelihood (ICL)} \label{app:icl}

\begin{figure}[!htb]
    \centering
    \subfigure[1978--1981.]{
        \includegraphics[scale=0.25]{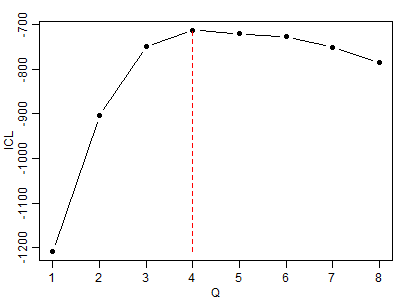}
    }
    \subfigure[1982--1984.]{
        \includegraphics[scale=0.25]{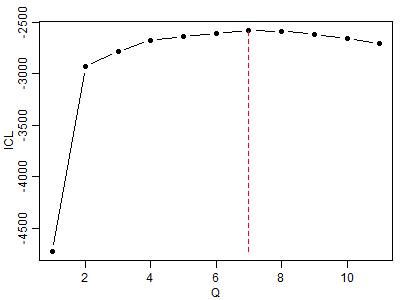}
    }
    \subfigure[1985--1987.]{
        \includegraphics[scale=0.25]{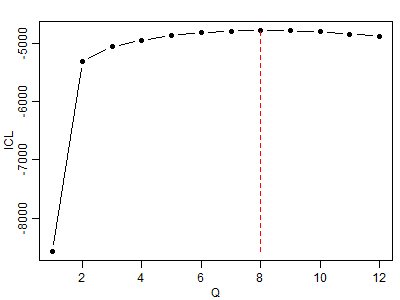}
    }
    \subfigure[1988--1990.]{
        \includegraphics[scale=0.25]{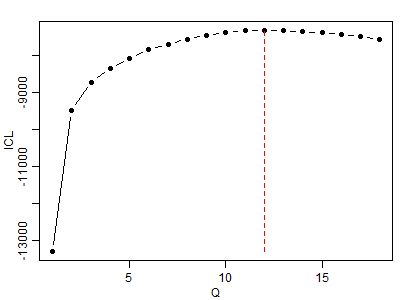}
    }
    \subfigure[1991--1993.]{
        \includegraphics[scale=0.25]{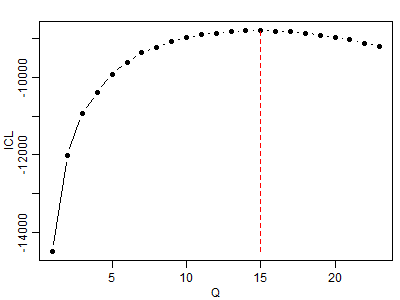}
    }
    \subfigure[1994--1996.]{
        \includegraphics[scale=0.25]{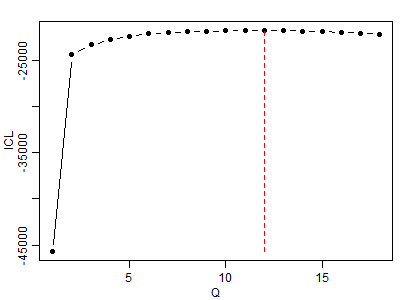}
    }
    \subfigure[1997--1999.]{
        \includegraphics[scale=0.25]{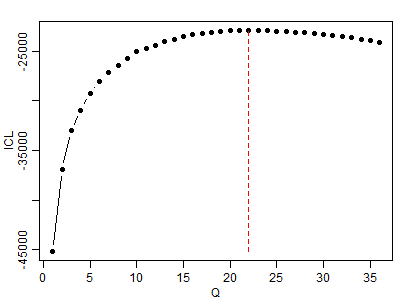}
    }
    \subfigure[2000--2001.]{
        \includegraphics[scale=0.25]{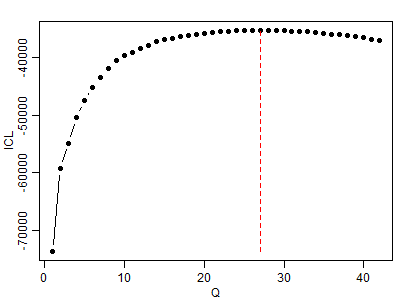}
    }
    \subfigure[2002--2004.]{
        \includegraphics[scale=0.25]{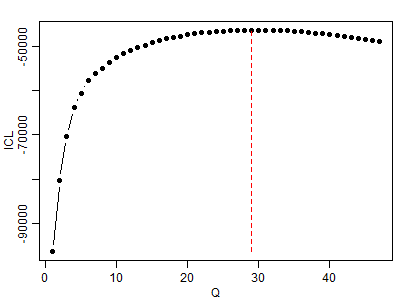}
    }
    \subfigure[2005--2007.]{
        \includegraphics[scale=0.25]{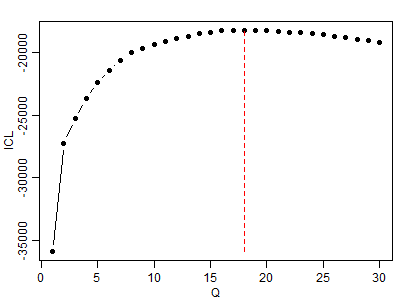}
    }
    \caption{Integration classification likelihood criterion as a function of the number of communities in the municipality networks.}
    \label{fig:ICLs_municipalities_combined}
\end{figure}

\end{document}